\documentclass[a4paper,12pt]{article}
\pdfoutput=1
\usepackage{epsfig}
\usepackage{amssymb}
\usepackage{amsfonts}
\usepackage{amsmath}
\usepackage{euscript}
\usepackage{verbatim}
\usepackage{latexsym}
\usepackage{graphicx}
\usepackage{caption}
\usepackage{float}
\usepackage{xcolor}
\usepackage{mathabx}
\usepackage{subcaption}
\usepackage[colorlinks=true]{hyperref}
\usepackage{cleveref}

\newif\ifdtup

\catcode`\@=11

\@addtoreset{equation}{section}

\def\@normalsize{\@setsize\normalsize{15pt}\xiipt\@xiipt
\abovedisplayskip 14pt plus3pt minus3pt%
\belowdisplayskip \abovedisplayskip
\abovedisplayshortskip \z@ plus3pt%
\belowdisplayshortskip 7pt plus3.5pt minus0pt}

\def\small{\@setsize\small{13.6pt}\xipt\@xipt
\abovedisplayskip 13pt plus3pt minus3pt%
\belowdisplayskip \abovedisplayskip
\abovedisplayshortskip \z@ plus3pt%
\belowdisplayshortskip 7pt plus3.5pt minus0pt
\def\@listi{\parsep 4.5pt plus 2pt minus 1pt
     \itemsep \parsep
     \topsep 9pt plus 3pt minus 3pt}}

\relax

\catcode`@=12

\catcode`\@=11

\def\section{\@startsection{section}{1}{\z@}{3.5ex plus 1ex minus
   .2ex}{2.3ex plus .2ex}{\large\bf}}

\def\SymBoxes#1#2#3#4{\newdimen\un@t \un@t#3%
\raisebox{#1}{\rule{#2\un@t}{#4}\hskip-#2\un@t
\@tempdimb\un@t \advance\@tempdimb by-#4\@tempcntb#2\relax%
\@whilenum{\@tempcntb>0}\do{
\rule{#4}{\un@t}\hskip\@tempdimb \advance\@tempcntb by\m@ne}%
\hskip-#2\un@t \rule[\un@t]{#2\un@t}{#4}%
\rule[\un@t]{#4}{#4}\hskip-#4
\rule{#4}{\un@t}}\hskip-#4}                

\begin{document}

\newcommand{\beq}{\begin{equation}}
\newcommand{\eeq}{\end{equation}}
\newcommand{\bea}{\begin{eqnarray}}
\newcommand{\eea}{\end{eqnarray}}
\newcommand{\beas}{\begin{eqnarray*}}
\newcommand{\eeas}{\end{eqnarray*}}
\newcommand{\defi}{\stackrel{\rm def}{=}}
\newcommand{\non}{\nonumber}
\newcommand{\bquo}{\begin{quote}}
\newcommand{\enqu}{\end{quote}}
\renewcommand{\(}{\begin{equation}}
\renewcommand{\)}{\end{equation}}
\def \eqn#1#2{\begin{equation}#2\label{#1}\end{equation}}

\def\e{\epsilon}
\def\IZ{{\mathbb Z}}
\def\IR{{\mathbb R}}
\def\IC{{\mathbb C}}
\def\IQ{{\mathbb Q}}
\def\de{\partial}
\def\Tr{ \hbox{\rm Tr}}
\def\H{ \hbox{\rm H}}
\def\HE{ \hbox{$\rm H^{even}$}}
\def\HO{ \hbox{$\rm H^{odd}$}}
\def\K{ \hbox{\rm K}}
\def\Im{ \hbox{\rm Im}}
\def\Ker{ \hbox{\rm Ker}}
\def\const{\hbox {\rm const.}}
\def\o{\over}
\def\im{\hbox{\rm Im}}
\def\re{\hbox{\rm Re}}
\def\bra{\langle}\def\ket{\rangle}
\def\Arg{\hbox {\rm Arg}}
\def\Re{\hbox {\rm Re}}
\def\Im{\hbox {\rm Im}}
\def\exo{\hbox {\rm exp}}
\def\diag{\hbox{\rm diag}}
\def\longvert{{\rule[-2mm]{0.1mm}{7mm}}\,}
\def\a{\alpha}
\def\dag{{}^{\dagger}}
\def\tq{{\widetilde q}}
\def\p{{}^{\prime}}
\def\W{W}
\def\N{{\cal N}}
\def\hsp{,\hspace{.7cm}}

\def\br{\nonumber}
\def\IZ{{\mathbb Z}}
\def\IR{{\mathbb R}}
\def\IC{{\mathbb C}}
\def\IQ{{\mathbb Q}}
\def\IP{{\mathbb P}}
\def \eqn#1#2{\begin{equation}#2\label{#1}\end{equation}}

\newcommand{\C}{\ensuremath{\mathbb C}}
\newcommand{\Z}{\ensuremath{\mathbb Z}}
\newcommand{\R}{\ensuremath{\mathbb R}}
\newcommand{\rp}{\ensuremath{\mathbb {RP}}}
\newcommand{\cp}{\ensuremath{\mathbb {CP}}}
\newcommand{\vac}{\ensuremath{|0\rangle}}
\newcommand{\vact}{\ensuremath{|00\rangle}                    }
\newcommand{\oc}{\ensuremath{\overline{c}}}
\newcommand{\psizero}{\psi_{0}}
\newcommand{\phizero}{\phi_{0}}
\newcommand{\hzero}{h_{0}}
\newcommand{\psiin}{\psi_{\rh}}
\newcommand{\phiin}{\phi_{\rh}}
\newcommand{\hin}{h_{\rh}}
\newcommand{\rh}{r_{h}}
\newcommand{\rb}{r_{b}}
\newcommand{\psibnd}{\psi_{0}^{b}}
\newcommand{\psibndp}{\psi_{1}^{b}}
\newcommand{\phibnd}{\phi_{0}^{b}}
\newcommand{\phibndp}{\phi_{1}^{b}}
\newcommand{\gbnd}{g_{0}^{b}}
\newcommand{\hbnd}{h_{0}^{b}}
\newcommand{\zh}{z_{h}}
\newcommand{\zb}{z_{b}}
\newcommand{\man}{\mathcal{M}}
\newcommand{\hbr}{\bar{h}}
\newcommand{\tbr}{\bar{t}}
\newcommand{\ee}[1]{\text{e}^{#1}}
\newcommand{\ii}{\text{i}}
\newcommand{\dd}{\text{d}}
\newcommand{\blue}[1]{\textcolor{purple}{#1}}

\begin{titlepage}
\begin{flushright}
TIFR/TH/20-4
\end{flushright}
\bigskip
\def\thefootnote{\fnsymbol{footnote}}

\begin{center}
{\Large
{\boldmath On Chebyshev Wells:\\
	Periods, Deformations, and Resurgence \\ \vspace{0.1in} 
}
}
\end{center}

\begin{center}
	Madhusudhan Raman$ ^{a} $\footnote{\texttt{madhur@theory.tifr.res.in}}, and  P. N. Bala Subramanian$^b$\footnote{\texttt{pnbalasubramanian@gmail.com}}\footnote{Current Affiliation: National Institute of Technology Calicut,	NIT Campus P.O 673 601,	Kozhikode, Kerala, India}
	\vspace{0.1in}

\end{center}

\renewcommand{\thefootnote}{\arabic{footnote}}

\begin{center}

$^a$ {Department of Theoretical Physics,\\ Tata Institute of Fundamental Research\\
	Homi Bhabha Road, Navy Nagar,\\ Colaba, Mumbai 400 005,\\ Maharashtra, India}
\\ \vspace{.1in}

$^b$ {Institute of Mathematical Sciences,\\ Homi Bhabha National Institute (HBNI)\\
	IV Cross Road, C.~I.~T.~Campus,\\ Taramani, Chennai 600 113,\\ Tamil Nadu, India}\\
\vspace{0.2in}
\end{center}
\noindent
\begin{center} {\bf Abstract} \end{center}
We study the geometry and mechanics (both classical and quantum) of potential wells described by squares of Chebyshev polynomials. We show that in a small neighbourhood of the locus cut out by them in the space of hyperelliptic curves, these systems exhibit low-orders/low-orders resurgence, where perturbative fluctuations about the vacuum determine perturbative fluctuations about non-perturbative saddles.

\vspace{1.6 cm}
\vfill
\end{titlepage}

\setcounter{footnote}{0}
\section{Introduction}

\subsection*{Periods}

When classical and quantum systems are studied via the lens of spectral curves, differential forms on them, and their integrals along cycles, surprising connections to more sophisticated lines of inquiry such as $ \mathcal{N} = 2 $ supersymmetric gauge theories \cite{Seiberg:1994rs}, topological strings on local Calabi-Yau geometries \cite{Katz:1996fh}, matrix models \cite{Akemann:1996zr,Dijkgraaf:2002fc,Dijkgraaf:2002vw,Dijkgraaf:2009pc}, and integrable systems \cite{Nekrasov:2009rc,Aganagic:2003qj,Gorsky:1995zq} are uncovered. 

There are now well-understood connections relating each of these lines of inquiry, for example the relation between topological strings and matrix models \cite{Bouchard:2007ys}. Another example of this fruitful interplay is \cite{Codesido:2016dld}, where it was shown that the holomorphic anomaly equations, which are usually discussed in the context of the refined topological string free energy, govern the WKB periods of one-dimensional quantum mechanical systems. Indeed, the relation between topological string theories and gauge theories on the one hand and quantum mechanical systems on the other has been well-studied \cite{Mironov:2009uv,Mironov:2009dv,He:2010xa,Maruyoshi:2010iu,Marshakov:2010fx,Krefl:2013bsa,Krefl:2014nfa,Kashani-Poor:2015pca,Ashok:2016yxz,Krefl:2016svj}. Leveraging identifications between spectral curves, problems in supersymmetric gauge theories with Seiberg-Witten curves can be mapped onto quantum mechanics problems, as was done in \cite{Grassi:2018spf}. Most recently, Argyres-Douglas points in the Coulomb branch of supersymmetric gauge theories with higher-rank gauge groups were studied by \cite{Ito:2020lyu} using techniques similar to the ones we will use. Finally, the equivalence between spectral determinants of Sturm-Liouville problems in the theory of ordinary differential equations, and the Baxter $ T $-$ Q $ relations in quantum integrable models has led to the formulation of the ODE/IM correspondence \cite{Dorey:1998pt}, which has also found applications in the study of resurgence \cite{Ito:2018eon}, supersymmetric gauge theories \cite{Fioravanti:2019vxi,Fioravanti:2019awr}, and string theory \cite{Komatsu:2019xzz,Dorey:2019ngq}.

Many of these studies have relied on the simple nature of geometry at genus-$ 1 $. Our goal in this paper will be to study quantum mechanical potentials corresponding to \emph{bona fide} higher-genus Riemann surfaces. Briefly, we will associate to our quantum mechanical system a (hyper-)elliptic curve of genus-$ g $ and study its quantum mechanics within the all-orders WKB framework \cite{Dunham32}. Classical and quantum WKB actions are treated as differential forms associated to the various cycles, which correspond in turn to classically allowed and forbidden regions as the case may be. This framework is clearly laid out in \cite{Fischbach:2018yiu} and we will follow it closely in our paper. An additional bonus is that this framework is naturally suited to the study of higher-genus $ (g>1) $ curves.

We focus in particular on differential operators that relate quantum WKB forms to their classical counterparts. The existence of these operators has consequences for the resurgent structure of the corresponding quantum systems, a discussion of which we now turn to.

\subsection*{Resurgence}

Resurgent asymptotics is the study of a complex web of relations that binds together perturbative and non-perturbative effects in quantum mechanical systems, both finite- and infinite-dimensional. Qualitatively, these relations come in two broad classes. The first --- large-orders/low-orders --- is expected to be a generic feature of quantum mechanical systems, and finds that the large-order growth of perturbation theory about one saddle is related to the low-order behavior of perturbation theory about another saddle \cite{Jentschura:2004jg,ZinnJustin:2004I,ZinnJustin:2004II,ZinnJustin:2004III}. The second --- low-orders/low-orders --- is believed to be less generic, and finds that information regarding the low-order behavior of perturbation theory about all saddles is contained in the low-order behavior of perturbation theory about the vacuum saddle \cite{Hoe82,Alvarez00Exponential,Alvarez00Uniform,Dunne:2013ada,Dunne:2014bca,Misumi:2015dua,Dunne:2016qix,Dunne:2016jsr}. This paper will concern itself with the latter form of resurgence in an infinite family of quantum mechanical potentials described by squares of Chebyshev polynomials and indexed by an integer $ m $, first discussed in \cite{Basar:2017hpr}. We will refer to these systems collectively as Chebyshev wells. They generalise the well-studied cubic and quartic oscillators.

At the classical level, these potentials enjoy an enhanced symmetry that ensures that period integrals of the classical action in any two valleys (respectively, peaks) are proportional to each other. Classically, and for arbitrary genus, each Chebyshev well traces out a (hyper)-elliptic curve that behaves as if it were a genus-$ 1 $ system, i.e.~it can be characterised by a single independent complex structure modulus. This lone complex structure modulus is in turn associated to a Hecke group H$ (m) $, a discrete subgroup of SL$(2,\mathbb{R})$. Consequently, we show in this paper that observables like periods and frequencies can be ``resummed'' into automorphic forms of Hecke groups, in a manner that is more reminiscent of similar resummations in supersymmetric gauge theories. This allows us to succinctly summarise many distinct formulas in the literature and write down compact and unified expressions valid for all Chebyshev wells.

For genus-$ 1 $ Chebyshev wells, the corresponding quantum theories are well-behaved; the proportionality of periods is preserved by quantum corrections. This raises the possibility of relating classical and quantum periods by differential operators, as was done in \cite{Basar:2017hpr}. 

For genus $ g > 1 $, the enhanced symmetry that confers upon classical periods and dual periods these nice properties is broken by quantum corrections. Consequently, the proportionality of periods is broken once quantum corrections are taken into account, and this in turn precludes the possibility of writing down differential operators that relate quantum periods to their classical counterparts. 

This symmetry breaking induced by quantum corrections is not wholly unfamiliar, and an analogy with supersymmetric gauge theories may be helpful. Equivariant localisation of $ \mathcal{N} = 2 $ supersymmetric gauge theories is effected in an $ \Omega $-background parametrised by two complex numbers $ \left(\epsilon_1,\epsilon_2\right) $, and instanton contributions to observables are first computed in the $ \Omega $-deformed theory, after which the deformation parameters are tuned to zero. The Nekrasov-Shatashvili limit sends only one of these (say, $ \epsilon_2$) to zero. It has been shown that the classical integrable system underlying the Seiberg-Witten theory is quantized in the Nekrasov-Shatashvili limit, with $ \epsilon_1 $ playing the role of Planck's constant \cite{Nekrasov:2009rc}. In the study of special vacua in supersymmetric gauge theories, it was observed that turning on $ \Omega $-deformations in the Nekrasov-Shatashvili limit breaks the symmetries of the special vacuum \cite{Ashok:2015cba}.

We will circumvent this obstruction to constructing differential operators by introducing further deformations within the broader geometrical framework we referred to earlier, with the ultimate goal of demonstrating that \emph{all} higher-genus Chebyshev wells exhibit the low-orders/low-orders resurgence property that is enjoyed more transparently by their genus-$ 1 $ counterparts.

\subsection*{Deformations}

The classical proportionality of the periods is a signature that the underlying geometry of the quantum mechanical system is singular. Indeed, similar geometries have arisen in the study of special vacua in supersymmetric gauge theories \cite{Ashok:2015cba,Ashok:2016oyh} and also in relation to replica surfaces in conformal field theories \cite{Calabrese:2009ez,Cardy:2017qhl}. We start by deforming our quantum mechanical system away from these special loci in the space of hyperelliptic curves. This has the effect of explicitly breaking the symmetry that forces the classical periods to be proportional to each other. The classical system then no longer behaves as if it were effectively of genus-$ 1 $.

The usefulness of deformations in teasing out the resurgent structure of quantum systems has been noted in the literature. For example, in \cite{Dunne:2016jsr} it is found that the ground state energy of supersymmetric quantum systems being zero to all orders in perturbation theory is best understood as the end result of a cancellation between two divergent series. This structure is uncovered by softly breaking supersymmetry via the introduction of a deformation parameter. Similarly, in \cite{Dorigoni:2017smz,Dorigoni:2019kux} supersymmetric theories are studied, where perturbative expansions are generally truncated and therefore the methods of resurgent analysis cannot be straightforwardly applied. To circumvent this obstruction, a supersymmetry-breaking deformation is introduced, which renders the perturbative expansions asymptotic. Finally, in \cite{Kozcaz:2016wvy} the importance of complex saddles and hidden topological angles in the study of quasi-exactly solvable systems is uncovered once again using deformations.

Of course, we will require that the deformation, parametrised by $ \eta $, is small. We insist on this requirement to ensure that the roots of the deformed potential are still on the real line, since if $ \eta $ is sufficiently large, roots can pair off and go into the complex plane, making the quantum mechanical interpretation of the corresponding period integrals difficult. Further, in the interest of continuity, we will also insist that the deformation doesn't change the genus of the spectral curve.

In this deformed curve, and using the techniques in \cite{Fischbach:2018yiu}, we show that differential operators that relate classical WKB forms to their quantum counterparts can in fact be constructed. The precise obstruction to constructing the requisite differential operators for higher-genus curves in the undeformed theory is clarified in the $ \eta \rightarrow 0 $ limit, and concomitantly, we use the $ \eta $-deformation to shed light on why it is possible to construct differential operators for genus-$ 1 $ potentials in the first place.

We now summarise the contents of the paper.

\subsection*{Outline}
In Section \ref{sec:Mechanics}, we associate to each quantum mechanical system under consideration a curve, introduce various observables like periods and frequencies, and introduce the all-orders WKB approximation. 

In Section \ref{sec:Curves} we study the geometry of these curves, and describe how to compute Picard-Fuchs differential equations (whose solutions are periods) and differential operators, closely following \cite{Fischbach:2018yiu}. In Section \ref{sec:GenusOne}, by way of example, we study in detail a genus-$ 1 $ system. This also gives us an opportunity to review the results of \cite{Basar:2017hpr}. 

In Section \ref{sec:GenusTwo} we study a genus-$ 2 $ system, first classically, where we show that the underlying Hecke symmetry can be used to write down universal expressions for classical observables. We then see that when quantum mechanical corrections are taken into account, the periods are no longer proportional to each other. This presents an obstruction to constructing differential operators that relate classical and quantum periods.

Finally, in Section \ref{sec:Deformations}, in order to circumvent this obstruction, we introduce the $ \eta $-deformation and revisit the genus-$ 2 $ example. We show that the requisite differential operators can be constructed for the deformed theory that arrange themselves into a Laurent series in $ \eta $. When acting on the deformed classical periods, we show that the $ \eta \rightarrow 0 $ limit is well-defined. As a result, the quantum periods of the undeformed theory can in fact be written as differential operators acting on the deformed classical periods. We also see in this section that the $ \eta $-deformation sheds light on why there is no such obstruction at genus-$ 1 $. 

Appendices \ref{app:PFEs} and \ref{app:DifferentialOperators} contain longer expressions for deformed Picard-Fuchs equations and differential operators respectively.

\subsection*{Acknowledgments}
This paper is dedicated to the memory of our fellow physicist, collaborator, and friend Koutha Vamsi Pavan Kumar.
	
We are grateful to Sujay Ashok, Arghya Chattopadhyay, Dileep Jatkar, Chethan Krishnan, and Hossein Movasati for discussions, and to G\"o\c{c}ke Ba\c{s}ar and Gerald Dunne for very helpful correspondence. MR acknowledges support from the Infosys Endowment for Research into the Quantum Structure of Spacetime.

\section{Mechanics}
\label{sec:Mechanics}

\subsection{Classical}
We consider a family of quantum mechanical potentials called Chebyshev wells that are indexed by an integer $ m \in \mathbb{Z}$ and $ m \geq 3 $, so we write
\begin{equation}\label{eq:VChebyshev}
V(x) = T_{m/2}^2(x) \ ,
\end{equation}
where $ T_{n}(\cos \theta) = \cos (n \theta) $, a Chebyshev polynomial of the first kind. Classically, particles with an energy $ \xi $ and momentum $ p $ find themselves bound by energy conservation to satisfy
\begin{equation}\label{key}
p^2 +V(x) = \xi \ ,
\end{equation}
where without loss of generality, we have chosen the mass of the particle to be $ 1/2 $. Much of the following discussion in this section is true quite generally, i.e.~it is not restricted to just the Chebyshev wells but arbitrary polynomial potentials.

In phase space, coordinatised by the canonically conjugate variables $ (x,p) \in \mathbb{C}^2 $, the above equation defines a family of hyperelliptic curve parametrised by $ \xi \in \hat{\mathbb{C}} $. This is more clearly visible if we rearrange the terms to write:
\begin{equation}\label{eq:CXi}
C_\xi : \quad p^2 =  \xi - V(x) \ .
\end{equation}
That is, our mechanical system is a hyperelliptic fibration over $ \hat{\mathbb{C}} $, the Riemann sphere parametrised by $ \xi $. We will often speak of the genus of a quantum mechanical system, by which we mean the genus of the corresponding hyperelliptic curve. For a polynomial of degree $ d $, the genus of the corresponding hyperelliptic curve is $ \lfloor \frac{d-1}{2} \rfloor $, where $ \lfloor \cdot \rfloor $ denotes the floor function. We'll have more to say about the geometry of this mechanical system in the following section.

For a specific value of the energy $ \xi \in [0,1] $, the quantum mechanical system is divided up into classically allowed and classically disallowed regions, which we will refer to as valleys and peaks respectively. The end-points $ \lbrace x_c \rbrace $ of these regions are determined by the condition
\begin{equation}\label{eq:TurningPoint}
p(x_c) = 0 \ .
\end{equation}
That is, classical turning points are defined as those points where the momentum of the particle vanishes.

To each valley, we can associate a period defined by
\begin{equation}\label{eq:ClassicalAction}
\widecheck{a}^{(0)}_i(\xi) = \oint_{A_i} \dd x \, p(x) \ , \\
\end{equation}
where $ A_i $ is an interval along the real line corresponding to the $ i^{\text{th}} $ valley, flanked by classical turning points. Similarly, we can associate to each peak a dual period defined by
\begin{equation}\label{eq:ClassicalDualAction}
\widehat{a}^{(0)}_i(\xi) = \oint_{B_i} \dd x \, p(x) \ , \\
\end{equation}
where $ B_i $ is an interval along the real line corresponding to the $ i^{\text{th}} $ peak. Once again, these intervals are flanked by classical turning points. The choice of notation for periods and dual periods is to remind the reader that periods $ (\widecheck{a}) $ are evaluated across valleys, while dual periods $ (\widehat{a}) $ are evaluated under peaks. The superscript $ (0) $ is to indicate the the observables are classical, i.e.~they contribute at $ O(\hbar^0) $.

These periods and dual periods correspond to oscillatory and evanescent motions across valleys and under peaks respectively. Other observables can be determined in terms of these periods as well. For example, one can define the frequencies and dual frequencies associated to these wells as
\begin{align}\label{key}
\begin{split} 
\widecheck{\omega}_i^{(0)}(\xi) &= \frac{\dd}{\dd \xi} \widecheck{a}_i^{(0)}(\xi) \ , \\
\widehat{\omega}_i^{(0)}(\xi) &= \frac{\dd}{\dd \xi} \widehat{a}_i^{(0)}(\xi) \ .
\end{split} 
\end{align}
For definiteness, we will restrict our attention to periods in this paper --- it will become clear that our techniques can equally well be applied to other observables such as frequencies. In fact, since our analyses will study not the periods themselves but $ 1 $-forms like $ p \, \dd x $ and their quantum analogues, by construction our results apply equally well to dual periods. The (classical) periods and dual periods are useful quantities to study when considering not only the classical mechanics of these wells, but also their quantum mechanical properties within the all-orders WKB framework, which we now turn to.

\subsection{Quantum}
The Schr\"odinger equation governing the dynamics of the corresponding quantum mechanical systems is
\begin{equation}
\hbar^{2} \psi^{\prime \prime}(x)+p^{2} \psi(x)=0 \ ,
\end{equation}
where $ p $ is given by \eqref{eq:CXi} and $ \psi(x) $ is the wavefunction of the particle. We study the quantum mechanics of these Chebyshev wells using a sophisticated version of the familiar WKB ansatz, and our presentation will follow the discussion in \cite{Codesido:2016dld}. Start with the ansatz
\begin{equation}
\psi(x)=\exp \left[\frac{\mathrm{i}}{\hbar} \int^{x} \mathrm{d} x^{\prime}\, Q\left(x^{\prime}\right) \right] \ ,
\end{equation}
where we assume that the action $ Q(x) $ admits a formal power series expansion in $ \hbar $ as
\begin{equation}
Q(x)=\sum_{n=0}^{\infty} Q_{n}(x) \hbar^{n} \ .
\end{equation}
When this all-orders WKB ansatz is fed into the Schr\"odinger equation, it implies a set of relation that determine the actions $ Q_n(x) $ recursively. It is easy to check that while the classical action $ Q_0(x) = p(x) $, the successive quantum actions are determined by the recursion relation
\begin{equation}\label{eq:QRecursion}
2 Q_{0}(x) Q_{n+1}(x)=\mathrm{i} Q_{n}'(x)-\sum_{k=1}^{n} Q_{k}(x) Q_{n+1-k}(x) \ .
\end{equation}

At this stage, we split the action into odd and even powers of $ \hbar $. Let 
\begin{equation}
Q(x)=Q_{\mathrm{odd}}(x)+Q_{\text{even}}(x)
\end{equation}
with
\begin{equation}
Q_{\text{even}}(x)=\sum_{n=0}^{\infty} Q_{2 n}(x) \hbar^{2 n} \ .
\end{equation}
Then the odd and even parts of the action are related as
\begin{equation}
Q_{\mathrm{odd}}(x)=\frac{\mathrm{i} \hbar}{2} \frac{\dd}{\dd x} \log Q_{\text{even}}(x) \ ,
\end{equation}
and so the all-orders WKB wavefunction is
\begin{equation}
\psi(x)=\frac{1}{\sqrt{Q_{\text{even}}(x)}} \exp \left[\frac{\mathrm{i}}{\hbar} \int^{x} \mathrm{d} x^{\prime} \, Q_{\text{even}}\left(x^{\prime}\right) \right] \ .
\end{equation}

The quantum actions $ Q_{2n>0} $ represent quantum corrections to the classical action $ Q_0 $, and so we can define quantum periods using them. To each valley, we can associate an infinite tower of quantum actions indexed by the superscript $ (2n) $ and defined by
\begin{equation}\label{eq:QuantumAction}
\widecheck{a}^{(2n)}_i(\xi) = \oint_{A_i} \dd x \, Q_{2n}(x) \ , \\
\end{equation}
where $ A_i $ is an interval along the real line corresponding to the $ i^{\text{th}} $ valley. Similarly, we can associate to each peak an infinite tower of quantum dual actions indexed by the superscript $ (2n) $ and defined by
\begin{equation}\label{eq:QuantumDualAction}
\widehat{a}^{(2n)}_i(\xi) = \oint_{B_i} \dd x \, Q_{2n}(x) \ , \\
\end{equation}
where $ B_i $ is an interval along the real line corresponding to the $ i^{\text{th}} $ peak. The classical and quantum actions together give formal expressions for quantum periods and quantum dual periods. Suppressing the valley and peak indices for a moment, we write
\begin{align}
\begin{split} 
\widecheck{a}(\xi, \hbar) &= \sum_{n=0}^{\infty} \hbar^{2 n} \, \widecheck{a}^{(2n)}(\xi) \ , \\
\widehat{a}(\xi, \hbar) &= \sum_{n=0}^{\infty} \hbar^{2 n} \, \widehat{a}^{(2n)}(\xi) \ .
\end{split} 
\end{align}
These quantum periods and quantum dual periods encode quantum corrections to oscillatory and evanescent motions in the valleys and peaks respectively.

\section{Curves}
\label{sec:Curves}

\subsection{Geometry}
In the previous section we assigned a genus $ g $ to the hyperelliptic curve $ C_\xi $ under consideration. This assignment may be pictured as two copies of the Riemann sphere $ \hat{\mathbb{C}} $ glued together along the branch cuts that connect the $ d $ classical turning points defined by \eqref{eq:TurningPoint}. We will introduce Picard-Fuchs differential equations and certain differential operators, and our discussion in this section will closely follow \cite{Fischbach:2018yiu}.

There are $ 2g $ independent $ 1 $-cycles on this geometry. The half of these that straddle classically allowed regions are what we referred to as $ A $-cycles in \eqref{eq:ClassicalAction}, while the other half that straddle classically forbidden regions are what we referred to as $ B $-cycles in \eqref{eq:ClassicalDualAction}. For example, a genus-$ 1 $ curve has one $ A $-cycle and one $ B $-cycle, and the integral of the classical action (a $ 1 $-form) $ Q_0 = p\,\dd x $ along these cycles yields the classical period and classical dual period respectively.

A choice of potential cuts out a slice in the space of hyperelliptic curves. We can think of motion of this slice in the space as parametrised by the energy $ \xi $, while motion within the slice for a fixed $ \xi $ is effected by M\"obius transformations of $ x $, since any two hyperelliptic curves are equivalent as long as their respective branch points are related by these fractional linear transformations. As we tune $ \xi $, it is easy to imagine these classical turning points merging, implying in turn the vanishing of the corresponding classical actions or dual actions. Points at which this happens are singular, and the curve $ C_\xi $ is said to degenerate at these points --- this corresponds to $ 1 $-cycles pinching off. 

This degeneration of hyperelliptic curves happens when the discriminant $ \Delta $ of the hyperelliptic curve, defined in terms of the roots $ \lbrace x_i \rbrace $ of the equation $ p^2 = 0 $ as
\begin{equation}\label{key}
\Delta=c \prod_{i<j}\left(x_{i}-x_{j}\right)^{2} \ ,
\end{equation}
vanishes. While for $ d\geq 5 $ there are no closed form expressions for the $ x_i $ via the Abel-Ruffini theorem, in our case the discriminant can always be written in terms of the critical points $ \lbrace x_{i}^{(c)} \rbrace $ of the potential $ V $ as
\begin{equation}\label{key}
\Delta(\xi)=\prod_{i=1}^{d-1}\left(\xi-V\left(x_{i}^{(c)}\right)\right) \ ,
\end{equation}
where it is explicitly seen to be a function of the single free parameter $ \xi $ in the curve, i.e.~the energy.

The geometrical picture that emerges is that of a hyperelliptic fibration of $ \hat{\mathbb{C}} $ minus the discriminant locus, so we identify the base space of our fibration as 
\begin{equation}
\hat{\mathbb{C}} \, \backslash \left(\lbrace \Delta = 0 \rbrace \cup \lbrace \infty \rbrace \right) \ .
\end{equation}
To each point on this base manifold, we associate a smooth hyperelliptic curve $ C_\xi $. Our discussion of $ 1 $-cycles (being associated to the first homology group) may be dualised to a statement about the first cohomology group, i.e.~the space of $ 1 $-forms at each point, which are sewn together to form a complex vector bundle over the base space. 

We will approach the computation of (classical and quantum) periods and dual periods via local solutions to the Picard-Fuchs equations, which we turn to next.

\subsection{Picard-Fuchs Equations}
As we saw in the previous section, the first cohomology groups associated to $ C_\xi $ at each point over the base space form a complex vector bundle. The periods and their derivatives will produce local sections on this vector bundle, and finite dimensionality implies that there will be a relation between them, so the differential equation that encapsulates the linear dependence will be at most of order $ 2g $. This differential equation is called the Picard-Fuchs equation. In the following, we will outline the method to construct these operators, given the period (or the corresponding action).

We'd like to solve the recursion \eqref{eq:QRecursion}, and to do so we introduce the following decomposition of the quantum actions
\begin{align}
\label{eq:QnGeneralForm}
Q_{n} = \dfrac{q_{n}}{p^{3n-1}} \ ,
\end{align}
where the polynomial $ q_{n} $ is obtained from the recursion relation
\begin{equation}
\label{eq:pnRecursionRelation}
q_{n+1}=\frac{\mathrm{i}}{4}\left(2 q_{n}^{\prime}(\xi-V)+(3 n-1) q_{n} V^{\prime}\right)-\frac{1}{2} \sum_{k=1}^{n} q_{k}\,\, q_{n+1-k} \ ,
\end{equation}
supplemented with boundary conditions 
\begin{equation}
\label{eq:pnRecursionBC}
q_{0}=1 \quad \text{and} \quad q_{1}=-\frac{\mathrm{i}}{4} V^{\prime} \ .
\end{equation}

The Picard-Fuchs equation $ \mathfrak{L}^{(n)} $ associated to the quantum action $ Q_{n} $, satisfies the relation
\begin{align}
\mathfrak{L}^{(n)} \left[ \oint \dd x \, Q_{n} \right] = 0 \ ,
\end{align}
or equivalently, the Picard-Fuchs equation annihilates the quantum action up to total derivatives,
\begin{align}
\mathfrak{L}^{(n)} \, Q_{n} = \dfrac{\dd}{\dd x}\bigg(\cdots\bigg).
\end{align}

In order to find the Picard-Fuchs equation satisfied by the quantum action $ Q_{n} $, following \cite{Fischbach:2018yiu}, we start with the ansatz
\begin{equation}
\label{eq:PFEGeneral}
\sum_{i=0}^{r} f_{i}(\xi) \frac{\dd^{i}}{\dd \xi^{i}}\left(\frac{q_{n}}{p^{3 n-1}}\right)=\sum_{k=0}^{k_{\max }} \frac{\dd}{\dd x}\alpha_{k}(\xi) \left(\frac{x^{k}}{p^{3 n-3+2 r}}\right) \ ,
\end{equation}
for some $ r \leq 2g$ and $ k_{\text{max}} $, where we are required to solve for the coefficients $ \alpha_k(\xi) $ and $ f_i(\xi) $. Let us take a moment to explain, operationally, what the above equation is doing. The left-hand side of the above equation with derivatives with respect to $ \xi $ is an ansatz for an order-$ r $ differential equation with arbitrary coefficients $ f_i(\xi) $ that must be fixed. The right-hand side of the above equation with derivatives with respect to $ x $ are total derivatives that will not affect the evaluation of period integrals. Our goal will be to use this indifference of period integrals to total derivatives to remove as many monomials as possible. Requiring that the coefficients of whatever remaining monomials vanish will fix the $ f_i(\xi) $. The Picard-Fuchs operator we finally obtain is 
\begin{align}
\mathfrak{L}^{(n)} = \sum_{i=0}^{r}f_{i}(\xi) \dfrac{\dd^i}{\dd \xi^i}.
\end{align}

The regular solutions around $ \xi = 0 $ are obtained via the Frobenius method, which generates the fundamental system of solutions $ \lbrace \Pi_{(i)} \rbrace $ to the Picard-Fuchs equation. The periods are then given by linear combinations of the fundamental system of solutions
\begin{align}
\widecheck{a}_{i}^{(n)} = \sum c_{i} \, \Pi_{(i)} \ ,
\end{align}
and the coefficients are fixed by comparing with perturbative calculations.

\subsection{Differential Operators}
The linear dependence that we used to construct Picard-Fuchs differential equations can also be leveraged to compute differential operators that act on classical actions to derive quantum actions. It is important to keep in mind that the statements we are making in this section pertain to the actions $ Q_{2n} $ rather than the periods $ \widecheck{a}^{(2n)} $ or dual periods $ \widehat{a}^{(2n)} $. Consequently, the notion of equality, in this case between a quantum action and a differential operator acting on the classical action, is only up to total derivatives. 

We start with an ansatz
\begin{equation}
\mathfrak{D}^{(2 n)} Q_{0} =\left[\sum_{i=0}^{r} d_{i}^{(2 n)}(\xi) \frac{\dd^{i}}{\dd \xi^{i}}\right] Q_{0} \ ,
\end{equation}
where $ d_i^{(2n)}(\xi) $ are unknown functions of $ \xi $ and $ r \leq 2g-1 $. We'd like this combination to be equal to $ Q_{2n} $. Note that these are components of $ 1 $-forms, i.e.~the objects of interest are the forms $ Q_{2n}\,\dd x $, and we are interested ultimately in periods that integrate these $ 1 $-forms over cycles of the genus-$ g $ hyperelliptic curve. We will consequently require that this equality is only up to total derivatives since by Stokes' theorem the latter do not affect the periods. 

Operationally, this is achieved in much the same way that the Picard-Fuchs equations were derived. We'd like to solve the equation
\begin{equation}
Q_{2n} - \mathfrak{D}^{(2 n)} Q_{0} = \sum_{\ell = 0}^{\ell_{\text{max}}} \beta_{\ell}(\xi) \frac{\dd}{\dd x} \left(\frac{x^\ell}{p^{6n-3}}\right) \ ,
\end{equation} 
with some $ \ell_{\text{max}} $ for the coefficients $ d_{i}^{(2n)}(\xi) $ and $ \beta_{\ell}(\xi) $. We proceed just as before, by requiring that the coefficients of all monomials are zero, and for an appropriate choice of $ \ell_{\text{max}} $ this system of equations may be possible to solve. When all these coefficients are solved for, we are guaranteed that
\begin{equation}
Q_{2n} \simeq \mathfrak{D}^{(2n)}Q_0 \ ,
\end{equation}
where $ \simeq $ indicates equality up to total derivatives.

For genus $ g>1 $ potentials of the form \eqref{eq:VChebyshev}, it turns out that such differential operators are not possible to compute, i.e.~there is no choice of $ \ell_{\text{max}} $ for which all the coefficients $ d_i^{(2n)}(\xi) $ and $ \beta_\ell(\xi) $ can be computed. Indeed, as we will see in specific examples, the Picard-Fuchs equation corresponding to \emph{all} potentials labelled by $ m\in\mathbb{Z}_{\geq 3} $ are always second-order --- this means that classically, the systems behaves as if there were only one period and dual period. The inability to compute appropriate differential operators is an indication that this is a singular geometry --- although not the kind where cycles are pinching off! We will discuss this case in more detail in the following sections.

We now work out a well-understood example at genus-$ 1 $.

\section{Genus One}
\label{sec:GenusOne}
In the follow sections, we evaluate classical and quantum periods corresponding to a genus-$ 1 $ potential. We also compute the appropriate differential operators that relate them.

\subsection{Classical Periods}
\begin{figure}[h]
	\centering
	\includegraphics[width=0.75\textwidth]{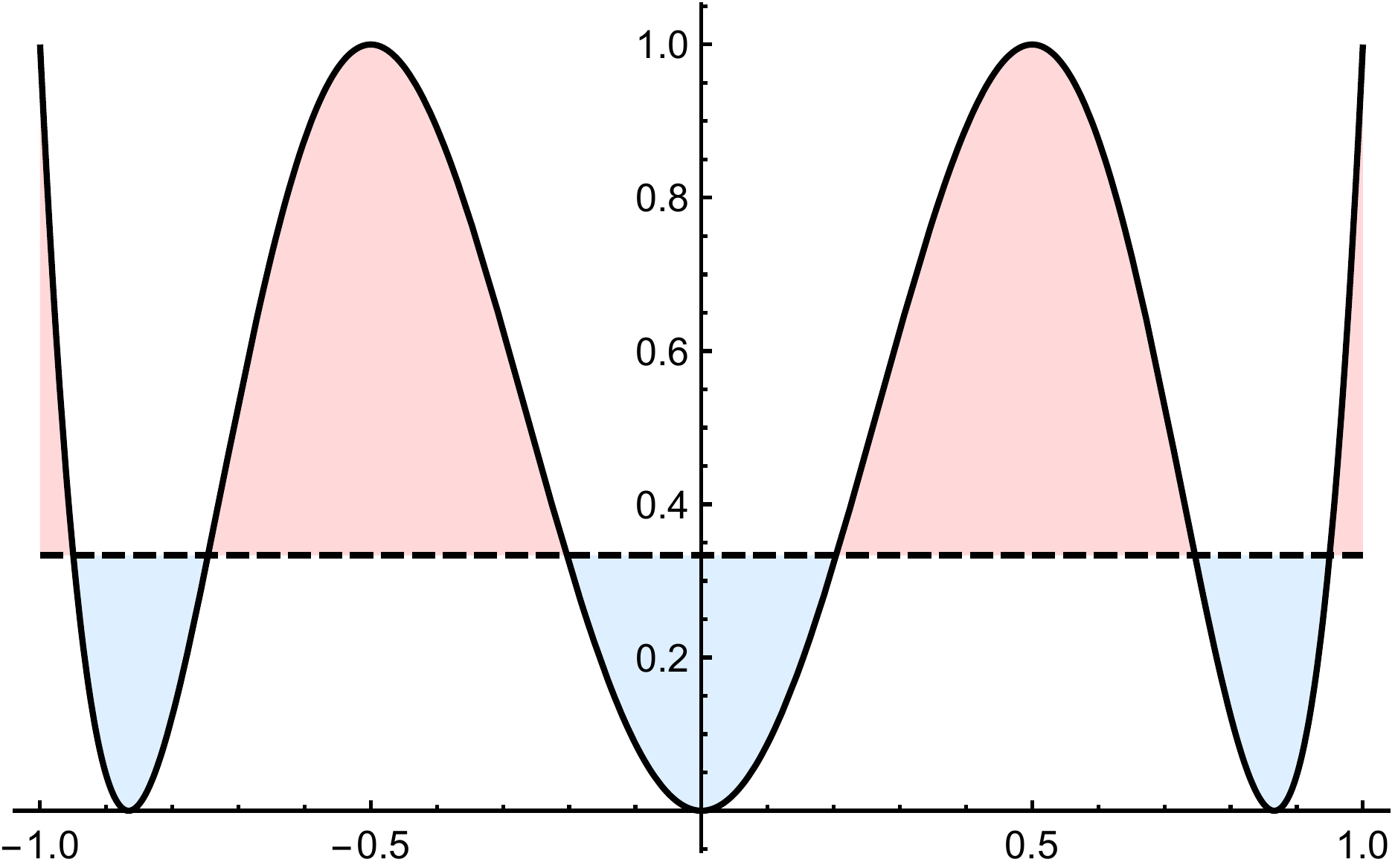}
	\caption{The solid line is the potential $ V = T_3^2(x) $, and the dashed line is some energy $ \xi $. The \textcolor{red}{red} regions are classically disallowed, while the \textcolor{blue}{blue} regions are classically allowed. By convention, we will label valleys and peaks starting from the left.}
	\label{fig:T3}
\end{figure}

Consider the case of $ m = 6 $, where the potential is of the form
\begin{equation}\label{key}
V(x) = 16 x^6-24 x^4+9 x^2 \ ,
\end{equation}
and is plotted in Figure \ref{fig:T3}. In this case there are \emph{prima facie} two independent valleys and two independent peaks. From simple contour deformation arguments, the period for the right-most valley around $ x = +\sqrt{3}/2 $ is not independent and is a determined in terms of the periods corresponding to the valleys around $ x = 0 $ and $ x = -\sqrt{3}/2 $. It is easy to see that if we choose coordinates $ z = x^2 $ then the above curve is of genus-$ 1 $, so the naive counting of two independent periods and dual periods is misleading.

\subsubsection{Classical Picard-Fuchs Equation}
The Picard-Fuchs differential equation corresponding to this potential is computed using \eqref{eq:PFEGeneral} and takes the form
\begin{equation}\label{key}
9 \xi (\xi-1)  \Pi''(\xi)+2 \Pi(\xi) = 0 \ .
\end{equation}
It is easy to see that the Picard-Fuchs equation knows about the genus-$ 1 $ nature of the curve, since the differential equation is only second-order. Linear combinations of the two independent solutions will make up the classical action and dual action. For definiteness, we will focus on the classical action.

Supplying the above equation with a Frobenius ansatz, we find that a solution to the above Picard-Fuchs differential equation is any linear combination of the form
\begin{equation}\label{key}
\Pi(\xi) = \alpha \left[\xi +\frac{1}{9}\xi ^2 + \cdots \right] + \beta \left[\left(1 - \frac{11}{9}\xi -\frac{5}{81}\xi^2 + \cdots \right) + \left(\frac{2}{9}\xi + \cdots \right)\log \xi \right]  \ ,
\end{equation}
with arbitrary $ \alpha,\beta\in\mathbb{C} $.

For future computations, it will be helpful to observe that on imposing the boundary conditions 
\begin{equation}\label{eq:PiBoundaryConditions}
\Pi(0)=0 \quad \text{and} \quad \Pi'(0) = 1 \implies \Pi(\xi) = \xi + O(\xi^2) \ ,
\end{equation}
we get the solution
\begin{align}\label{key}
\begin{split} 
\Pi(\xi) &= \xi +\frac{1}{9}\xi ^2+\frac{10}{243}\xi ^3 + \cdots \ , \\
&= \xi \ {}_2 F_1 \left(\frac{1}{3},\frac{2}{3},2;\xi \right) \ .
\end{split} 
\end{align}

\subsubsection{Perturbative Computation of Classical Periods}

Our expectation is that the classical period should be a solution to the Picard-Fuchs equation. The classical action is
\begin{equation}\label{key}
Q_0 = p \, \dd x \ .
\end{equation}
Let us now compute these period integrals perturbatively. We will not pay attention to overall factors multiplying period integral computations, and instead leave them implicit. That is, the results for the perturbative computation of the periods should be understood as up to overall multiplicative $ \xi $-independent normalisation factors. 

We expand the $ 1 $-form $ p \, \dd x $ about $ \xi = 0 $ and denote $ \underline{p} = p\vert_{\xi = 0} $ to get
\begin{equation}\label{eq:pdxExpansion}
\underline{p}\,\dd x + \frac{\dd x}{2\underline{p}} \xi - \frac{\dd x}{8\underline{p}^3} \xi^2 + \frac{\dd x}{16 \underline{p}^5} \xi^3 + \cdots \ .
\end{equation} 
We now integrate this expansion about the valley around $ x = -\sqrt{3}/2 $, and we get
\begin{align}\label{key}
\begin{split} 
\widecheck{a}_1^{(0)}(\xi) &=  \xi +\frac{1}{9}\xi ^2+\frac{10}{243}\xi ^3 + \cdots  \ , \\
&= \xi \ {}_2 F_1 \left(\frac{1}{3},\frac{2}{3},2;\xi \right) \ ,
\end{split} 
\end{align}
which matches the solution to the Picard-Fuchs equation, as expected. A similar computation can be done around the valley about $ x = 0 $ and we find that
\begin{align}\label{key}
\begin{split} 
\widecheck{a}_2^{(0)}(\xi) &=  \xi +\frac{1}{9}\xi ^2+\frac{10}{243}\xi ^3 + \cdots  \ , \\
&= \xi \ {}_2 F_1 \left(\frac{1}{3},\frac{2}{3},2;\xi \right) \ ,
\end{split} 
\end{align}
and so we see that the two \emph{prima facie} independent periods $ a_1^{(0)}(\xi) $ and $ a_2^{(0)}(\xi) $ are in fact proportional to each other, differing only by a $ \xi $-independent constant.\footnote{Since the two periods are identical, we will occasionally suppress the valley index.} Of course, this is not so surprising when we take into account the fact that the potential is really genus-$ 1 $. We now turn to a perturbative evaluation of the quantum periods but, before that, briefly address the question of dual periods and also the classical geometrical properties of this potential. 

\subsection{Geometry, Duality, and Modular Forms}
\label{sec:GDModular}
Just as we computed the classical periods perturbatively, with a little more work we can also compute the classical dual periods. For the potential in question, it was shown in \cite{Basar:2017hpr} that using the beautiful identity
\begin{equation}
_2 F_{1}\left(\frac{1}{p}, 1-\frac{1}{p}, 1 ; z\right)=\frac{2}{\pi} \int_{0}^{\arcsin \sqrt{z}} \dd \theta \, \frac{\cos \left(\left(\frac{2}{p}-1\right) \theta\right)}{\sqrt{z-\sin ^{2} \theta}}
\end{equation}
the classical dual period in either valley is computed to be 
\begin{equation}\label{key}
\widehat{a}^{(0)}(\xi) = (1-\xi) \ {}_2F_{1}\left( \frac{1}{3},\frac{2}{3},2;1-\xi \right) \ ,
\end{equation}
and together with the results of the previous section, we can define a modular parameter 
\begin{align}\label{eq:tauXiRelation}
\begin{split} 
\tau &= -\frac{\ii}{\sqrt{3}} \frac{\dd \widehat{a}^{(0)}/\dd\xi}{\dd \widecheck{a}^{(0)}/\dd\xi} \ , \\
&= \frac{\ii}{\sqrt{3}} \frac{{}_2F_{1}\left( \frac{1}{3},\frac{2}{3},1;1-\xi \right)}{{}_2F_{1}\left( \frac{1}{3},\frac{2}{3},1;\xi \right)} \ , 
\end{split}
\end{align}
whose associated modular transformations are generated by the following operations:
\begin{equation}\label{eq:STonTau}
T: \tau \rightarrow \tau + 1 \quad \text{and} \quad S: \tau \rightarrow -\frac{1}{3\tau} \ .
\end{equation}
As observed in \cite{Basar:2017hpr} this transformation property is precisely the transformation property satisfied by generators of the Hecke group H$ (6) $, a discrete subgroup of SL$ (2,\mathbb{R}) $. An element of H$ (6) $, more abstractly, is any word made up of the letters $ T $ and $ S $ circumscribed by the relations
\begin{equation}
S^2 = 1 \quad \text{and} \quad (ST)^6 = 1 \ .
\end{equation}
Furthermore, H$ (6) $ is an arithmetic Hecke group, meaning that it has finite index in SL$ (2,\mathbb{Z}) $ and consequently that its theory of modular forms corresponds to the theory of modular forms associated to the congruence subgroup $ \Gamma_{0}(3) \subset \text{SL}(2,\mathbb{Z})$.

A generalisation of the Jacobi inversion formula \cite{Berndt95,Cooper09} allows us to now express the periods in terms of $ \tau $, the modular parameter, instead of $ \xi $, the energy. This might seem like an odd thing to do, so let us first present an analogy with supersymmetric gauge theories which might be helpful. In \cite{Ashok:2015cba} non-perturbative corrections to the dual periods and the period matrix of an $ \mathcal{N} = 2 $ supersymmetric gauge theory with gauge group SU$ (3) $ and $ N_f = 6 $ fundamental flavours were computed first via equivariant localisation, which returns answers in terms of the ``bare coupling'' $ q_0 $. In this work and work that preceded it \cite{Huang:2011qx,Billo:2013fi,Huang:2013eja,Billo:2013jba}, it was consistently found that while the $ q_0 $-expansions were largely inscrutable, if one performs a non-perturbative redefinition of the coupling to a new ``renormalized coupling'' $ q $, the resulting $ q $-expansions neatly organise themselves into modular forms of either the modular group or one of its congruence subgroups, depending on the context. We will find that much the same is true of the change of variables from $ \xi $ to $ \tau $.

The specific form of the change of variables can be written in terms of the Dedekind $ \eta $-function 
\begin{align}
\eta(\tau) &= \ee{\ii\pi\tau/12} \prod_{n=1}^{\infty} \left(1-\ee{2\pi\ii\tau  n}\right) = q^{1/24} \prod_{n=1}^{\infty} \left(1-q^n\right) \ ,
\end{align}
where the elliptic nome is related to the modular parameter as $ q = \ee{2\pi\ii\tau} $. The appropriate Jacobi inversion formula is now given by
\begin{equation}
\label{eq:JacobiInversionH6}
\frac{\xi}{1-\xi} = 27 \left(\frac{\eta(3\tau)}{\eta(\tau)}\right)^{12} \ .
\end{equation}
This is a wholly remarkable formula. To see why, first recall the well-known functional equation satisfied by the Dedekind $ \eta $-function
\begin{equation}
\eta\left(-\frac{1}{\tau}\right) = \sqrt{\frac{\tau}{\ii}} \, \eta (\tau) \ .
\end{equation}
Now consider an $ S $-transformation as in \eqref{eq:STonTau} acting on the right hand side of the above equation, which sends it to its inverse. In order to preserve the above identification under $ S $-transformations, we are forced to induce an $ S $-action on $ \xi $:
\begin{equation}
\label{eq:xiSTransform}
S: \xi \rightarrow 1-\xi \ .
\end{equation}
Note that the Chebyshev wells have all their minima at $ \xi = 0 $ and all their maxima at $ \xi = 1 $. Indeed, that this ought to be true is evident from the definition of the modular parameter in \eqref{eq:tauXiRelation}. If $ \tau $ is likened to a coupling in a gauge theory, then the $ S $-transformation sends weak to strong coupling. Analogously, we might liken expansions near $ \xi = 0 $ to ``electric'' expansions, and expansions near $ \xi = 1 $ might be likened to ``magnetic'' expansions in Seiberg-Witten theory. The classical theory, then, has a duality symmetry that relates classical periods and dual periods by the simple replacement in \eqref{eq:xiSTransform}.

We intend to use the above formula to re-write $ \xi $-expansions as $ q $-expansions. Solving for $ \xi $ gives us
\begin{align}
\begin{split} 
\xi &= \frac{27 \left(\frac{\eta(3\tau)}{\eta(\tau)}\right)^{12}}{1+27 \left(\frac{\eta(3\tau)}{\eta(\tau)}\right)^{12}} \ , \\
&= 27q \left(1 - 15 q + 171 q^2 - 1679 q^3 + \cdots \right) \ .
\end{split} 
\end{align}

Various observables may now be cast as $ q $-expansions. For example, the classical period, instead of admitting a small-$ \xi $ expansion, can now be thought of as admitting a small-$ q $ expansion (alternatively, an expansion in the limit $ \tau \rightarrow \ii\infty $), the first few terms of which are
\begin{equation}
\widecheck{a}^{(0)}(\tau) = 27q \left(1 - 12 q + 111 q^2 - 908 q^3 + \cdots \right) \ .
\end{equation}
Remarkably, in \cite{Basar:2017hpr} the above $ q $-expansion was resummed in terms of the weight two and weight four Eisenstein series of SL$ (2,\mathbb{Z}) $
\begin{equation}
\begin{aligned}
E_2(\tau) &=1-24 \sum_{n=1}^{\infty} \frac{n q^{n}}{1-q^{n}} \ , \\
E_4(\tau) &=1+240 \sum_{n=1}^{\infty} \frac{n^{3} q^{n}}{1-q^{n}} \ ,
\end{aligned}
\end{equation}
as
\begin{equation}
\widecheck{a}^{(0)}(\tau) = \frac{3\sqrt{2}}{16} \frac{9\left(E_2^2(3\tau) - E_4(3\tau)\right)-\left(E_2^2(\tau) - E_4^2\right)}{\left(3E_2(3\tau) - E_2(\tau)\right)} \ .
\end{equation}

In fact, for each of the arithmetic Chebyshev wells (corresponding to $ m = 3,4,6,\infty $) the authors of \cite{Basar:2017hpr} were able to resum the classical periods and dual periods into various combinations of the Eisenstein series. However, their expressions are unsatisfactory for a couple of reasons. First, the expressions look different for each of the arithmetic Chebyshev wells, i.e.~there is no unified expression for these observables. Second, the non-arithmetic cases are not treated at all, despite possessing the same kinds of symmetries as the arithmetic cases at the classical level. 

We will address both these problems in Section \ref{sec:GDAutomorphic}, where we will bring to bear the our understanding of automorphic forms of Hecke groups to write down universal expressions for the classical periods.

\subsection{Quantum Periods}
The second quantum period, computed using \cref{eq:QnGeneralForm,eq:pnRecursionRelation,eq:pnRecursionBC}, is
\begin{equation}\label{key}
Q_2 = \frac{V''}{8p^3} + \frac{5\left(V'\right)^2}{32p^5} \ .
\end{equation}
As before, we expand this about $ \xi = 0 $ and evaluate this quantum period in the first valley, which we find to be
\begin{equation}\label{key}
\widecheck{a}^{(2)}_1(\xi) = 1 + \frac{20}{27} \xi + \frac{490}{729} \xi^2 + \frac{1400}{2187} \xi^3 + \cdots  \ .
\end{equation}
We can perform the same computation the second valley, and we find
\begin{equation}\label{key}
\widecheck{a}^{(2)}_2(\xi) = 1 + \frac{20}{27} \xi + \frac{490}{729} \xi^2 + \frac{1400}{2187} \xi^3 + \cdots \ .
\end{equation}
We have here perturbatively reproduced the conclusions of \cite{Basar:2017hpr} where it was found that the quantum periods continue to be proportional to each other.

On noting this striking property of genus-$ 1 $ systems, \cite{Basar:2017hpr} highlighted the interesting possibility that one could construct differential operators that allowed one to determine quantum periods from classical ones. Since all the periods are proportional to each other, i.e.~$ a_k^{(2n)} \propto a_{\ell}^{(2n)} $, we can dispense with the valley index. Then
\begin{equation}
\widecheck{a}^{(2n)} = \mathfrak{D}^{(2n)} \, \widecheck{a}^{(0)} \ ,
\end{equation}
where $ \mathfrak{D}^{(2n)} $ is a differential operator. For example, the first two differential operators are found to be
\begin{align}\label{key}
\mathfrak{D}^{(2)} &= \left(-\frac{2}{3 \xi}-\frac{4}{9 (\xi-1)}\right) + \frac{5}{3} \frac{\dd}{\dd\xi} \ , \\
\nonumber \mathfrak{D}^{(4)} &= \left(\frac{21}{5 \xi^3}+\frac{1}{5 \xi^2}+\frac{164}{405 (\xi-1)^2}-\frac{56}{45 (\xi-1)^3}+\frac{49}{81 \xi}-\frac{49}{81 (\xi-1)}\right) \\
&\quad + \left(-\frac{21}{5 \xi^2}+\frac{56}{45 (\xi-1)^2}+\frac{4}{15 \xi}-\frac{4}{15 (\xi-1)}\right) \frac{\dd}{\dd\xi} \ .
\end{align}

Further, since the same differential operators determine quantum corrections to dual periods --- remember that our construction of the differential operators made reference only to the forms at play, and not the period integrals themselves! --- they concluded that the arithmetic (genus-1) Chebyshev wells (with $ m \in \lbrace 3, 4, 6, \infty \rbrace$) were ``P/NP resurgent'', meaning that low-order quantum corrections to perturbative motion also determine low-order quantum corrections to non-perturbative motion.\footnote{Since resurgence refers to the relations between perturbative and non-perturbative physics, we find the term ``P/NP resurgence'' ambiguous, and instead will refer to this form of resurgence as ``low-orders/low-orders resurgence'' instead.} As we have discussed in the introduction, the goal of this note is to consider the higher-genus Chebyshev wells and see to what extent this form of resurgence exists. This is the subject of the following section.

\section{Higher Genus}
\label{sec:GenusTwo}
In this section, we will discuss the resurgent properties of higher-genus Chebyshev wells, starting with a discussion of the classical periods and subsequently the quantum periods and the construction of differential operators that relate classical and quantum periods. For definiteness, we will focus on a genus-$ 2 $ Chebyshev well.

\subsection{Classical Periods}
\begin{figure}[h]
	\centering
	\includegraphics[width=0.75\textwidth]{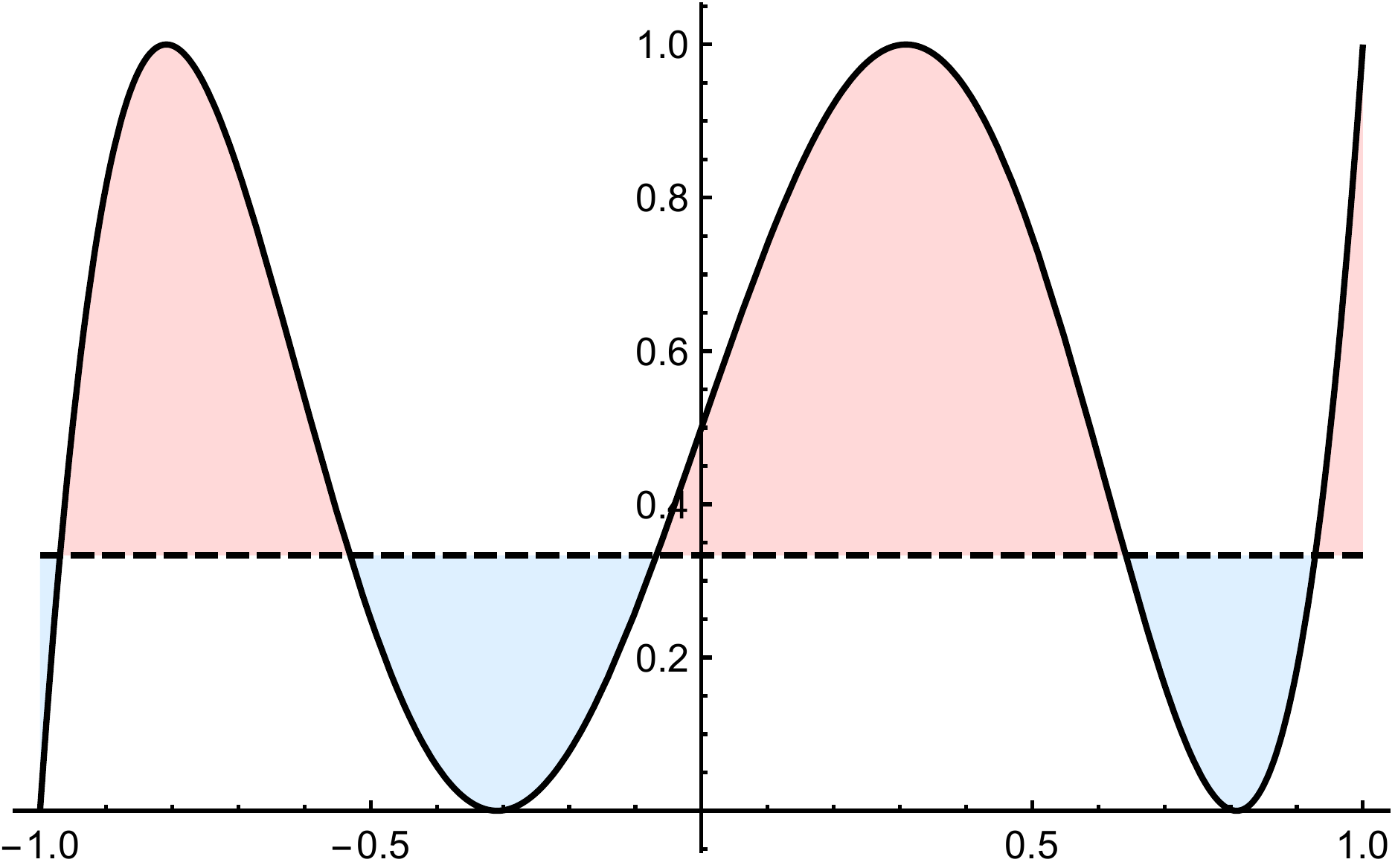}
	\caption{The solid line is the potential $ V = T_{5/2}^2(x) $, and the dashed line is some energy $ \xi $. The \textcolor{red}{red} regions are classically disallowed, while the \textcolor{blue}{blue} regions are classically allowed. Once again, by convention, we will label valleys and peaks starting from the left --- in this case there are \emph{prima facie} two independent valleys and two independent peaks.}
	\label{fig:T52}
\end{figure}

Consider the case of $ m = 5 $, where the potential is of the form
\begin{equation}\label{key}
V(x) = 8 x^5-10 x^3+\frac{5}{2}x+\frac{1}{2} \ ,
\end{equation}
and is plotted in Figure \ref{fig:T52}. It is a \emph{bona fide} genus-$ 2 $ potential, i.e.~there is no change of variables that will make the above potential have degree $ \leq 4 $. However, as we will see, this curve's classical period satisfies a second-order differential equation, implying that classically it behaves as if it were effectively of genus-$ 1 $.

\subsubsection{Classical Picard-Fuchs Equation}

The Picard-Fuchs equation corresponding to this curve, computed using \eqref{eq:PFEGeneral}, is
\begin{equation}\label{key}
100 \xi (\xi-1) \Pi''(\xi)+21 \Pi(\xi) = 0 \ .
\end{equation}
Of particular note, it is a second-order differential equation, with only two linearly independent solutions. This is surprising, and indeed, one would expect a generic genus-$ 2 $ potential to have a Picard-Fuchs differential equation of fourth-order, so as to accommodate two independent periods and two dual periods. Given that this reduction of the order of the classical Picard-Fuchs equation is a property of all potentials given by \eqref{eq:VChebyshev}, we conclude that they all (classically) behave as if they were ``effectively'' of genus-$ 1 $, meaning that all their periods and dual periods are the same up to $ \xi $-independent normalisations.

Equivalently, when thinking of the period matrix of a generic genus-$ 2 $ hyperelliptic curve, one would expect there to be three independent complex structure moduli, organised into a symmetric $ 2\times 2 $ matrix. However, since the classical Picard-Fuchs equation is second-order, implying in turn that there is effectively only one independent period and one independent dual period, at the level of the period matrix there is only one complex structure modulus. Intuitively, we might say that this potential corresponds to a special, singular slicing of the space of hyperelliptic curves where there are no degenerations (i.e.~no cycles pinching off), but where the \emph{prima facie} independent periods (respectively, dual periods) are all proportional to each other. These surfaces have arisen before, for example in the study of supersymmetric gauge theories with gauge group SU$ (N) $ and $ N_f = 2N $ fundamental flavours \cite{Minahan:1995er,Argyres:1998bn,Ashok:2016oyh}. Here, the classical vacuum expectation values of the adjoint scalar in the $ \mathcal{N} = 2 $ vector multiplet were be arranged in such a way that they only differ from each other by roots of unity, endowing the vacuum with a $ \mathbb{Z}_N $ symmetry --- this was called the special vacuum. Similar surfaces have also arisen in the study of replica surfaces \cite{Calabrese:2009ez,Cardy:2017qhl} relevant for entanglement entropy calculations in two-dimensional conformal field theories. 

It is important to keep in mind that all of this is only true when quantum corrections are ignored. Once quantum corrections are taken into account, this degeneracy is lifted --- the analogous statement in the supersymmetric gauge theory context would be that the $ \Omega $-deformations generically break the $ \mathbb{Z}_N $ symmetry associated to the special vacuum. We will see all this in the quantum mechanical system we are studying more explicitly in the following sections.

Before we conclude this section, we note that imposing the boundary conditions \eqref{eq:PiBoundaryConditions} as before, we find the following solution to the Picard-Fuchs equation:
\begin{align}\label{key}
\begin{split} 
\Pi(\xi) &= \xi +\frac{21}{200}\xi ^2+\frac{1547}{40000}\xi ^3 + \cdots \ , \\
&= \xi \ {}_2 F_1 \left(\frac{3}{10},\frac{7}{10},2;\xi \right) \ .
\end{split} 
\end{align}

\subsubsection{Perturbative Computation of Classical Periods}

Let us now compute these period integrals perturbatively. We expand the period $ p \, \dd x $ about $ \xi = 0 $ as in \eqref{eq:pdxExpansion}, integrate this expansion term by term about the valley around $ x = \frac{1}{4}\left(1-\sqrt{5}\right) $, and we get
\begin{align}\label{key}
\begin{split} 
\widecheck{a}_1^{(0)}(\xi) &= \xi +\frac{21}{200}\xi ^2+\frac{1547}{40000}\xi ^3 + O(\xi^4)  \ , \\
&= \xi \ {}_2 F_1 \left(\frac{3}{10},\frac{7}{10},2;\xi \right) \ ,
\end{split} 
\end{align}
which matches the solution to the Picard-Fuchs equation, as expected. A similar computation can be done around the valley about $ x = \frac{1}{4}\left(1+\sqrt{5}\right) $ and we find that
\begin{align}\label{key}
\begin{split} 
\widecheck{a}_2^{(0)}(\xi) &= \xi +\frac{21}{200}\xi ^2+\frac{1547}{40000}\xi ^3 + O(\xi^4)  \ , \\
&= \xi \ {}_2 F_1 \left(\frac{3}{10},\frac{7}{10},2;\xi \right) \ ,
\end{split} 
\end{align}
and so we see that the two \emph{prima facie} independent periods $ a_1^{(0)}(\xi) $ and $ a_2^{(0)}(\xi) $ are in fact proportional to each other, differing only by a $ \xi $-independent normalisation. Note that this is an exceptional property and is by no means a generic fact of genus-$ 2 $ (or even higher-genus) systems. It is also consistent with the fact that the classical Picard-Fuchs equation is of second order.

We conclude from this demonstration that the classical theory has an enhanced symmetry, which we characterise geometrically in the following section.

\subsection{Geometry, Duality, and Automorphic Forms}
\label{sec:GDAutomorphic}
Just as we encountered expressions for the dual periods in the genus-$ 1 $ case, with the genus-$ 2 $ potential as well we can show with a little more work that the classical dual period in either valley is 
\begin{equation}\label{key}
\widehat{a}^{(0)}(\xi) = (1-\xi) \ {}_2F_{1}\left( \frac{3}{10},\frac{7}{10},2;1-\xi \right) \ ,
\end{equation}
and in fact for any $ m $, the classical periods and dual periods are of the form
\begin{align}
\begin{split} 
\widecheck{a}^{(0)}(\xi) &= \xi \ {}_2F_{1}\left( \frac{1}{2} - \frac{1}{m},\frac{1}{2} + \frac{1}{m},2;\xi \right) \ , \\
\widehat{a}^{(0)}(\xi) &= (1-\xi) \ {}_2F_{1}\left( \frac{1}{2} - \frac{1}{m},\frac{1}{2} + \frac{1}{m},2;1-\xi \right) \ .
\end{split} 
\end{align}
In the remainder of this section, we will write down unified expressions for arbitrary $ m $, thereby circumventing the problems raised in Section \ref{sec:GDModular}. In the interest of uniformity, we will restrict ourselves to finite $ m $. As in the genus-$ 1 $ example, we can define a modular parameter 
\begin{align}\label{key}
\begin{split} 
\tau &= -\frac{\ii}{2\cos\left(\frac{\pi}{m}\right)} \frac{\dd \widehat{a}^{(0)}/\dd\xi}{\dd \widecheck{a}^{(0)}/\dd\xi} \ , \\
&= \frac{\ii}{2\cos\left(\frac{\pi}{m}\right)} \frac{{}_2F_{1}\left( \frac{1}{2} - \frac{1}{m},\frac{1}{2} + \frac{1}{m},1;1-\xi \right)}{{}_2F_{1}\left( \frac{1}{2} - \frac{1}{m},\frac{1}{2} + \frac{1}{m},1;\xi \right)} \ , 
\end{split} 
\end{align}
whose associated modular transformations are generated by the following operations:
\begin{equation}\label{eq:TSActionHecke}
T: \tau \rightarrow \tau + 1 \quad \text{and} \quad S: \tau \rightarrow -\frac{1}{\lambda_m \tau} \ ,
\end{equation}
with
\begin{equation}
\lambda_m = 4\cos^2\left(\frac{\pi}{m}\right) \ .
\end{equation}
This transformation property is precisely the transformation property satisfied by generators of the Hecke group H$ (m) $, a discrete subgroup of SL$ (2,\mathbb{R}) $.\footnote{The Hecke groups H$ (m) $ are sometimes defined after rescaling the $ \tau $ variable, in which case the $ T $- and $ S $-transformations act as
	\begin{equation*}
	T: \tau \rightarrow \tau + \sqrt{\lambda_m} \quad \text{and} \quad S: \tau \rightarrow -\frac{1}{\tau} \ .
	\end{equation*}
} More abstractly, an element of H$ (m) $ is any word made up of the letters $ T $ and $ S $ circumscribed by the relations
\begin{equation}
S^2 = 1 \quad \text{and} \quad (ST)^m = 1 \ .
\end{equation}
Generically, H$ (m) $ is not arithmetic, i.e.~it does not have finite index in SL$ (2,\mathbb{Z}) $ and consequently, its theory of modular (rather, automorphic) forms is not related to some congruence subgroup of $ \text{SL}(2,\mathbb{Z})$. However, there is now a well-developed theory of automorphic forms associated to these Hecke groups \cite{Doran:2013npa,Raman:2018owg,Ashok:2018myo}, and we refer the reader to these papers for a more explicit realisation of the objects we will invoke in this section.

A generalisation of the Jacobi inversion formula appropriate to these groups was presented in \cite{Raman:2018owg}, and is given by
\begin{equation}
\label{eq:HeckeJacobiInversion}
\frac{\xi}{1-\xi} = \frac{\sqrt{j_m(\tau)} -\sqrt{j_m(\tau) - d_m} }{\sqrt{j_m(\tau)} + \sqrt{j_m(\tau) - d_m}} \ ,
\end{equation}
where $ j_m(\tau) $ is an analogue of the Klein $ j $-invariant of SL$ (2,\mathbb{Z}) $ associated to the Hecke group H$ (m) $, and $ d_m $ is the value of $ j_m(\tau) $ at the fixed point of the $ S $-action,
\begin{equation}
j_m\left(\frac{\mathrm{i}}{\sqrt{\lambda_{m}}}\right) = d_m \ ,
\end{equation}
in much the same way that the Klein $ j $-invariant has the property
\begin{equation}
j(\ii) = 1728 \ .
\end{equation}
As a check, from \cite{Zagier2002} we see that the analogue of the $ j $-invariant appropriate to the congruence subgroup $ \Gamma_0(3) $ (which, as we saw in Section \ref{sec:GDModular} is related to H$ (6) $ as well) is
\begin{equation}
j_{6}(\tau)=\left[\left(\frac{\eta(\tau)}{\eta(3 \tau)}\right)^{6}+27\left(\frac{\eta(3 \tau)}{\eta(\tau)}\right)^{6}\right]^{2} \ .
\end{equation}
On plugging the above expression into \eqref{eq:HeckeJacobiInversion} with $ d_6 = 108 $, we recover the formula \eqref{eq:JacobiInversionH6}. In this way, \eqref{eq:HeckeJacobiInversion} generalises the Jacobi inversion formula to all Hecke groups, and when the corresponding Hecke group is arithmetic it reproduces the Jacobi inversion formulas corresponding to congruence subgroups of the modular group.

We might wonder how the duality transformations work, since $ j_m(\tau) $ are invariant under Hecke action. This was worked out first in the context of supersymmetric gauge theories in \cite{Ashok:2016oyh}: the monodromy of $ j_m(\tau) $ around the fixed point of the $ S $-action as given in \eqref{eq:TSActionHecke} is evaluated by sending 
\begin{equation}
\left(j_m\left(\tau\right)-d_{m}\right) \rightarrow \mathrm{e}^{2 \pi \mathrm{i}}\left(j_m\left(\tau\right)-d_{m}\right) \ ,
\end{equation}
which in turn inverts the right hand side of \eqref{eq:HeckeJacobiInversion}. The induced $ S $-action on $ \xi $ is once again given by
\begin{equation}
S: \xi \rightarrow 1-\xi \ .
\end{equation}
This story is reminiscent of S-duality in supersymmetric gauge theories, where monodromies around the points where the whole Coulomb branch is singular generate the S-duality group \cite{Argyres:1998bn}. For completeness, the $ T $-action leaves the identification invariant, as we would expect from the analogy with supersymmetric gauge theories.

Solving for $ \xi $ in \eqref{eq:HeckeJacobiInversion} gives us
\begin{align}
\xi &= \frac{1}{2} \left(1-\frac{\sqrt{j_m(\tau)-d_m}}{\sqrt{j_m(\tau)}}\right) \ .
\end{align}

Various observables may now be cast as $ q $-expansions, courtesy of the above expression. The strategy would be to start by computing the observable in a small-$ \xi $ expansion, then substituting $ \xi $ by the above $ q $-series, and then resumming the quantity into some combination of automorphic forms of H$ (m) $. For example,
the classical frequencies and periods for \emph{all} Chebyshev wells are given by the unified expressions
\begin{align}
\begin{split} 
\widecheck{\omega}^{(0)}(\tau) &= \left(E_4^{(m)}\right)^{1/4} \ , \\
\widecheck{a}^{(0)}(\tau) &= \frac{\left( E_2^{(m)}E_4^{(m)} - E_6^{(m)}\right)}{\left(E_4^{(m)}\right)^{5/4}} \ ,
\end{split} 
\end{align}
where $ E_{2}^{(m)} , E_{4}^{(m)} $ and $ E_{6}^{(m)} $ are Eisenstein series of weights two, four, and six corresponding to the Hecke groups H$ (m) $. Explicit $ q $-series, along with Ramanujan identities and the like are discussed in great detail in \cite{Ashok:2018myo}. Finally, the duality symmetry we have highlighted implies expressions for the dual frequencies and dual periods are determined by the $ S $-action on the frequencies and periods respectively. More generally, if the classical observable is expressed in terms of $ \tau $ or $ \xi $, the appropriate dual is obtained by sending 
\begin{equation}
\tau \rightarrow - \frac{1}{\lambda_m \tau} \quad \text{or} \quad \xi \rightarrow 1-\xi \ ,
\end{equation}
as the case may be.

\subsection{Quantum Periods}
In the same manner that we computed the classical periods as a perturbative expansion in $ \xi $, we can evaluate the quantum periods perturbatively. The first quantum period, computed about the valley around $ x = \frac{1}{4}\left(1-\sqrt{5}\right)$, is given by
\begin{align}
\label{eq:a12}
\widecheck{a}_{1}^{(2)} &= \frac{3 \sqrt{5}+\blue{49}}{80}+\frac{\left(627 \sqrt{5}+\blue{7745}\right) \xi }{16000} + \frac{77 \left(1653 \sqrt{5}+\blue{18679}\right) \xi ^2}{3200000}+ \cdots \ ,
\end{align}
and the quantum period evaluated about the valley around $ x = \frac{1}{4}\left(1+\sqrt{5}\right)$ is given by
\begin{align}
\label{eq:a22}
\widecheck{a}_{2}^{(2)} &= \frac{3 \sqrt{5}+\blue{55}}{80}+ \frac{\left(627 \sqrt{5}+ \blue{8999}\right) \xi }{16000} + \frac{77 \left(1653 \sqrt{5}+\blue{21985}\right) \xi ^2}{3200000}+ \cdots \ .
\end{align}
We have highlighted the differences in the above expressions. It is important to note that when we computed the classical periods, both $ \widecheck{a}_{1}^{(0)} $ and $  \widecheck{a}_{2}^{(0)} $ were normalized so that the coefficient of the leading $ O(\xi) $ term is unity. In computing $ \widecheck{a}_{1}^{(2)} $, we normalize it with the same factor used in the case of $ \widecheck{a}_{1}^{(0)} $, and $ \widecheck{a}_{2}^{(2)} $ is normalized with the same factor as $ \widecheck{a}_{2}^{(0)} $.

It is clear from the explicit expressions above that the symmetry that set the classical periods to be proportional to each other is broken when quantum corrections are taken into account. The quantum periods $ a_k^{(2)}(\xi) $ are annihilated by a Picard-Fuchs operator $ \mathfrak{L}^{(2)} $, which takes the form
\begin{align}
\begin{split} 
\mathfrak{L}^{(2)} &= 10000 (\xi -1)^2 \xi ^2 \frac{\dd^{4}}{\dd \xi^{4}}+60000 \left(2 \,\xi ^2-3 \, \xi +1\right) \xi  \frac{\dd^{3}}{\dd \xi^{3}}\\
& \quad \ +1000 \left(377 \, \xi ^2-377 \, \xi +60\right) \frac{\dd^{2}}{\dd \xi^{2}}+154000 (2 \, \xi -1) \frac{\dd}{\dd \xi}+33649 \ .
\end{split} 
\end{align}
In terms of the (regular) fundamental system of solutions to the above differential equation given by
\begin{align}
\begin{split} 
\Pi_{(1)} &= 1+\frac{17 \, \xi }{20} +\frac{32417 \, \xi ^2}{40000} + \frac{633633 \, \xi ^3}{800000}+\cdots \ , \\
\Pi_{(2)} &= \xi +\frac{77 \, \xi ^2}{60} +\frac{340417 \, \xi ^3}{240000} + \frac{35942599 \, \xi ^4}{24000000} +\cdots \ , 
\end{split} 
\end{align}
the quantum periods \cref{eq:a12,eq:a22} can be written as
\begin{align}
\begin{split} 
\widecheck{a}_{1}^{(2)} & = \frac{3 \sqrt{5}+49}{80} \,~ \Pi_{(1)} + \frac{117 \left(\sqrt{5}-5\right)}{16000} \,~ \Pi_{(2)} \ ,\\
\widecheck{a}_{2}^{(2)} & = \frac{3 \sqrt{5}+55}{80} \,~ \Pi_{(1)} + \frac{117 \left(\sqrt{5}-3\right)}{16000} \,~ \Pi_{(2)} \ .
\end{split} 
\end{align}
The absence of a differential operator that relates the classical and quantum periods is evident from the fact that the quantum periods are not proportional to each other. This is because such a differential operator would have to generate two different branches of solutions by acting on a single object, which is obviously impossible. We now supply a prescription to nevertheless compute such differential operators, not in the original theory, but in a deformation thereof. 

\section{Deformations}
\label{sec:Deformations}
In the previous section, we saw that due to the lack of proportionality between the quantum periods in a genus-$ 2 $ potential, it is not possible to construct a differential operator that determines quantum periods from their classical counterparts. In this section we perform the same kinds of computations we have been doing, except that we start with a deformed potential
\begin{equation}
V(x) = T_{m/2}^2(x) + \eta \, x \ ,
\end{equation}
where $ \eta $ is a small, positive real number. We then construct differential operators with respect to both free parameters $ (\xi,\eta) $ and compute all quantities in the $ \eta $-deformed theory, only taking the limit $ \eta \rightarrow 0 $ limit at the end. In constructing the differential operators as well as the Picard-Fuchs equations, we avoid the use of derivatives with respect to $ \eta $, so that taking the limit $ \eta\rightarrow 0 $ becomes easier.\footnote{As emphasised in \cite{Fischbach:2018yiu}, there is no obstruction in principle to doing this. In fact, the resulting differential operators are of lower order in both the $ \xi $- and $ \eta $-derivatives. However, the $ \eta \rightarrow 0 $ limit is more subtle in this case.} This excursion will not only allow us to construct differential operators for this deformed curve, but also shed light on why it was possible to construct differential operators for the genus-$ 1 $ system in the first place.

For small $ \eta $, the differential operators can be organised as a Laurent series of differential operators in $ \eta $. Further, the exact form of the differential operator will depend on the nature of the deformation chosen. Indeed, if one chooses a slightly different deformation, say by $ \eta \, x^2 $, the form of the deformed Picard-Fuchs equations and differential operators will be different. Our interest, however, is fundamentally in the quantum periods, so to this end the exact form of these differential operator is immaterial. One restriction we will make, however, is to always deform by a monomial that doesn't change the genus of the curve. This is reasonable from a geometrical point of view since we do not want the genus of the hyperelliptic curve to change in the $ \eta \rightarrow 0 $ limit, which we would like to arrange to be smooth.

\subsection{Genus Two, Deformed}
Let us start by first looking at the Picard-Fuchs equations satisfied by the classical periods (and dual periods) of the deformed theory. Keeping with the requirement of only including $ \xi $ derivatives, we compute the associated Picard-Fuchs equation for arbitrary $ \eta $, which can be expressed as a series in $ \eta $
\begin{align}
\label{eq:etaDeformedPFE}
\mathfrak{L}^{(0)} &= \sum_{k=0} \mathfrak{L}^{(0)}_{k} \eta^{k} .
\end{align}
For our purposes, we only keep track of the operators upto linear order in $ \eta $ --- our reasons for doing this will shortly become clear. (Detailed expressions for the $ \mathfrak{L}^{(0)}_0 $ and $ \mathfrak{L}^{(0)}_1 $ are given in Appendix \ref{app:PFEs}.) Given these differential equations, we now find the fundamental system of solutions for the above operator as a double series expansion in $ \xi $ and $ \eta $. Restricting our attention to regular expansions (no logarithmic terms) we find that the periods are a linear combination of the following branches of solutions:
\begin{align}
\begin{split} 
\Pi_{(1)} &= \left(1+ \frac{1503 \, \xi^2}{40000}+ \cdots \right) +\eta\,\dfrac{1881\, \xi^{2}}{3200000}\left(37+ \dfrac{79933 \, \xi}{2000}+ \cdots\right)+ O\left(\eta^{2}\right) \ ,\\
\Pi_{(2)} &= \left(\xi +\dfrac{21 \, \xi^2}{200}+ \dfrac{1547 \, \xi^3}{40000}+ \cdots\right) + O\left(\eta^{2}\right) \ ,\\
\Pi_{(3)} &=  \eta \left(1+ \dfrac{1503 \, \xi^2}{40000}+ \dfrac{56679 \, \xi^3}{2000000}+ \cdots\right) + O\left(\eta^{2}\right) \ ,\\
\Pi_{(4)} &=  \eta \left(\xi +\dfrac{21 \, \xi^2}{200}+ \dfrac{1547 \, \xi^3}{40000}+ \cdots\right) + O\left(\eta^{2}\right) \ .
\end{split} 
\end{align}
The classical periods satisfy the Picard-Fuchs equation $ \mathcal{L}^{(0)} \Pi = 0 $, which means that the classical periods associated to each well are a linear combination of the above branches of solutions that make up the fundamental system. In order to fix the coefficients in the linear combination, we evaluate the classical periods perturbatively as a double series in $ \xi $ and $\eta$. We find that the classical periods corresponding to the two valleys are given by  
\begin{align}
\begin{split} 
\widecheck{a}_{1,\eta}^{(0)} &= \widecheck{a}_{1,0}^{(0)} + \eta \, \widecheck{a}_{1,1}^{(0)} + \cdots \ , \\ 
& =\Pi_{(2)} + \dfrac{\left(-1+\sqrt{5}\right)}{4}\Pi_{(3)} +
\dfrac{9 \left(-1+\sqrt{5}\right)}{400}\Pi_{(4)} \ ,
\end{split} 
\end{align}
\begin{align}
\begin{split}  
\widecheck{a}_{2,\eta}^{(0)} &= \widecheck{a}_{2,0}^{(0)} + \eta \, \widecheck{a}_{2,1}^{(0)} + \cdots \ , \\
& =\Pi_{(2)} + \dfrac{\left(1+\sqrt{5}\right)}{4}\Pi_{(3)} +
\dfrac{9 \left(1+\sqrt{5}\right)}{400}\Pi_{(4)} \ .
\end{split} 
\end{align}
This notation is unfortunately a bit unwieldy, so let us take a moment to explain it. In $ \widecheck{a}^{(n)}_{k,\ell} $, the superscript $ (n) $ indicates the order in $ \hbar $ at which this action contributes, the subscript $ k $ indicates the valley, and the subscript $ \ell $ indicates the order in the $ \eta $ expansion. As we can see, the subleading corrections in the $ \eta $-expansion of the periods computed in the left and right wells do not match. We see that the $ \eta $-deformation explicitly breaks the symmetry which had forced the classical periods in both valleys to be proportional to each other. Also, note that we have normalized the classical periods in such a way that the coefficient of $ \eta^{0}\xi^1 $ is unity for both the wells. We will subsequently normalize every quantum period with the corresponding normalization factor. It can be thought of as an overall factor pulled out of the $ \hbar $-expansion of the full quantum period in each well.

It should be emphasised, however, that the above expansions are classical data corresponding to a deformed curve. The $ \eta $-deformation is essentially a trick that allows classical computations in the deformed curve to be related to quantum periods of the undeformed curve via a set of differential operators, so in fact the strategy we will now outline really does relate classical and quantum data. 

With that in mind, we turn now to the differential operators, which can in fact be constructed in the deformed theory, and expanded as a series in $ \eta $. We find schematically that the differential operators take the form
\begin{equation}
\label{eq:D2nGeneral}
\mathfrak{D}^{(2n)} = \frac{1}{\eta} \, \mathfrak{D}_{-1}^{(2n)} + \mathfrak{D}_{0}^{(2n)} + \eta \, \mathfrak{D}_{1}^{(2n)} + \cdots \ ,
\end{equation}
where each of the differential operators in the $ \eta $-expansion take the form
\begin{equation}\label{key}
\mathfrak{D}_{k}^{(2n)} = \sum_{i=0}^{r} d_{i}^{(2 n)}(\xi) \frac{\dd^{i}}{\dd \xi^{i}} \ .
\end{equation}
For example, the first couple of differential operators in the $ \eta $-expansion of $ \mathfrak{D}^{(2)} $ are
\begin{align}\label{key}
\mathfrak{D}_{-1}^{(2)} &= \frac{5}{\delta_{-1}}\sum_{i=0}^{r}  d_{i}^{(2)}\, \frac{\dd^{i}}{\dd \xi^{i}} \ ,
\end{align}
where
\begin{align}
\begin{split} 
\delta_{-1} & = 12(512 \, \xi^2-512\, \xi+175) \ , \\
d_{0}^{(2)} &=  -126 \, \xi + 63 \ , \\
d_{1}^{(2)} &=   84 (\xi -1) \xi \ , \\
d_{2}^{(2)} &=  100 \, \xi  \left(2 \, \xi ^2-3 \, \xi +1\right) \ , \\
d_{3}^{(2)} &= 400 (\xi -1)^2 \xi ^2  \ ,
\end{split} 
\end{align}
and \begin{align}\label{key}
\mathfrak{D}_{0}^{(2)} &= \frac{1}{\delta_{0}}\sum_{i=0}^{r}  h_{i}^{(2)}\, \frac{\dd^{i}}{\dd \xi^{i}} \ ,
\end{align}
where
\begin{align}
\delta_{0} & = 30(512 \, \xi^2-512\, \xi+175)^{2} \ , \nonumber \\
h_{0}^{(2)} &= -63 \left(21504 \, \xi ^3-32256 \, \xi ^2+21202 \, \xi -5225\right)\ , \nonumber \\
h_{1}^{(2)} &=   15 \left(538624 \, \xi ^4-1077248 \, \xi ^3+928524 \, \xi ^2-389900 \, \xi +67375\right) \ , \\
h_{2}^{(2)} &= 50 \left(436224 \, \xi ^5-1090560 \, \xi ^4+1258148 \, \xi ^3-796662 \, \xi ^2+271600 \, \xi -39375\right) \ , \nonumber \\
h_{3}^{(2)} &= 500 \, \xi  \left(2048 \, \xi ^5-6144 \, \xi ^4+10644 \, \xi ^3-11048 \, \xi ^2+5375 \, \xi -875\right)  \ . \nonumber 
\end{align}

Clearly, the limit $ \eta \rightarrow 0 $ is not well-defined since the expansion \eqref{eq:D2nGeneral} is naively divergent. This divergence, however, will prove crucial in what follows. Let us proceed with the above differential operator, always continuing to work in the deformed context.

Acting with $ \mathfrak{D}^{(2)} $ on $ \widecheck{a}_{1,\eta}^{(0)} $, we get 
\begin{align}
\begin{split} 
\widecheck{a}_{1,\eta}^{(2)} = \mathfrak{D}^{(2)} a_{1,\eta}^{(0)} &= \left(\frac{1}{\eta} \mathfrak{D}_{-1}^{(2)} + \mathfrak{D}_{0}^{(2)} + \eta \, \mathfrak{D}_{1}^{(2)} +\cdots \right) \left(\widecheck{a}_{1,0}^{(0)} + \eta \, \widecheck{a}_{1,1}^{(0)} + \cdots \right)~ , \\
&= \frac{1}{\eta} \, \mathfrak{D}_{-1}^{(2)} \widecheck{a}_{1,0}^{(0)} + \left(\mathfrak{D}_{0}^{(2)}\widecheck{a}_{1,0}^{(0)} + \mathfrak{D}_{-1}^{(2)}\widecheck{a}_{1,1}^{(0)}\right) + \cdots \ .
\end{split} 
\end{align}
There are two remarkable facts about the above expression. The first is that the coefficient of the $ \eta^{-1} $ piece vanishes identically, i.e.~the undeformed classical period $ \widecheck{a}_{1,0}^{(0)} $ is annihilated by the differential operator $ \mathfrak{D}_{-1}^{(2)} $. An identical argument holds for the other period as well. This fact is crucial for the well-definedness of the $ \eta \rightarrow 0 $ limit.

The second remarkable fact is that 
\begin{equation}
\widecheck{a}_{1,0}^{(2)} = \mathfrak{D}_{0}^{(2)}\, \widecheck{a}_{1,0}^{(0)} + \mathfrak{D}_{-1}^{(2)}\,\widecheck{a}_{1,1}^{(0)} \ ,
\end{equation}
or equivalently, that the first quantum period in the undeformed theory corresponding to the first valley is determined by differential operators acting on the $ \eta $-deformed classical periods. The same is true of the second well:
\begin{equation}
\widecheck{a}_{2,0}^{(2)} = \mathfrak{D}_{0}^{(2)}\, \widecheck{a}_{2,0}^{(0)} + \mathfrak{D}_{-1}^{(2)}\, \widecheck{a}_{2,1}^{(0)} \ .
\end{equation}

The next check would be to ensure that the quantum periods computed perturbatively match precisely with the quantum period derived via the differential operators acting on the deformed classical periods. We find that this is in fact the case, and the quantum periods computed in both ways take the same form:
\begin{align}
\begin{split} 
\widecheck{a}_{1,0}^{(2)} &= \frac{3 \sqrt{5}+49}{80}+\frac{\left(627 \sqrt{5}+7745\right) \xi }{16000} + \frac{77 \left(1653 \sqrt{5}+18679\right) \xi ^2}{3200000}+ \cdots \ ,\\
\widecheck{a}_{2,0}^{(2)} &= \frac{3 \sqrt{5}+55}{80}+ \frac{\left(627 \sqrt{5}+ 8999\right) \xi }{16000} + \frac{77 \left(1653 \sqrt{5}+21985\right) \xi ^2}{3200000}+ \cdots \ .
\end{split} 
\end{align}

Since the $ \eta \rightarrow 0 $ limit is well-defined courtesy of the vanishing coefficient of $ \eta^{-1} $, we can then conclude that while it is not possible to naively find differential operators that send classical periods to quantum periods, it is possible to do so when starting with the $ \eta $-deformed curve. We have checked that the same story holds for the quantum periods $ \widecheck{a}_{k}^{(4)} $ and the corresponding differential operator $ \mathfrak{D}^{(4)} $, and we expect it to hold for all higher quantum periods. (Expressions for the first couple of differential operators in the $ \eta $-expansion of $ \mathfrak{D}^{(4)} $ are given in Appendix \ref{app:DifferentialOperators}.) Further, while our illustration is in a genus-$ 2 $ example, we expect the $ \eta $-deformation to be more generally applicable for all higher-genus Chebyshev wells.

\subsection{Genus One, Revisited}
In light of the above deformation, let us revisit the genus-$ 1 $ example we had encountered in an earlier section. (One should make a deformation by $ \eta\, x^2 $ in this case, so as to preserve the genus of the original curve, which is really $ 1 $.)  There, the deformed Picard-Fuchs equations once again take the form
\begin{equation}
\mathfrak{L}^{(0)} = \mathfrak{L}_0^{(0)} + \eta \, \mathfrak{L}_1^{(0)} + \cdots \ .
\end{equation}
However, the differential operator now looks like
\begin{equation}
\mathfrak{D}^{(n)} = \mathfrak{D}_{0}^{(n)} + \eta \, \mathfrak{D}_{1}^{(n)} + \cdots \ .
\end{equation}
Observe that there is no piece proportional to $ \eta^{-1} $. This of course implies that one can smoothly take the $ \eta \rightarrow 0 $ limit --- indeed, not introduce it at all as was done in \cite{Basar:2017hpr} --- and actually find a differential operator that relates the classical and quantum periods.

\section{Conclusions}

In this paper, we have studied the classical and quantum mechanics of an infinite family of potential wells described by squares of Chebyshev polynomials. We have extended the results of previous studies in two complimentary directions.

Classically, we have leveraged the duality symmetry associated to each potential
\begin{equation}
T_{m/2}^2(x) \longleftrightarrow \text{H}(m) \ ,
\end{equation}
to resum observables like periods and frequencies into combinations of automorphic forms of Hecke groups. This allows for simple, unified expressions valid for all $ m $ and not just for those potentials whose spectral curves are genus-$ 1 $. 

Quantum mechanically, in order to sidestep the obstruction to constructing differential operators that relate classical and quantum periods on higher-genus Riemann surfaces, we introduced a deformation of the hyperelliptic curve. We observed that in the deformed curve, the requisite differential operators could in fact be constructed. Focusing on the limit $ \eta \rightarrow 0 $, we saw that the quantum periods of the undeformed curve were determined in terms of differential operators acting on the classical data associated to the $ \eta $-deformed curve. The correctness of this prescription was checked against perturbative calculations of the quantum periods. While the example we worked out $ (m=5) $ was at genus-$ 2 $, it is sufficiently generic and strongly suggests that the $ \eta $-deformation can be used to arrive at similar a conclusion for all Chebyshev wells. 

The implications of this construction for low-orders/low-orders resurgence are striking. The differential operators we have constructed relate the classical WKB forms to their quantum counterparts up to total derivatives, so it is true by construction that the same differential operators that relate classical to quantum period integrals (which control perturbative physics) also relate classical to quantum dual period integrals (which control non-perturbative physics). Since these differential operators can be constructed when the $ \eta $-deformation is turned on, and when $ \eta $ is small, we conclude from this that a small neighbourhood of the locus cut out by Chebyshev wells in the space of hyperelliptic curves exhibits low-orders/low-orders resurgence, greatly expanding the set of examples for which this form of resurgence has been established.

\appendix

\section{Picard-Fuchs Equations}
\label{app:PFEs}
The $ \eta $-deformed Picard-Fuchs equation can be written as an expansion in $ \eta $ as in \eqref{eq:etaDeformedPFE}, and the differential operators for the first couple of orders in the $ \eta $-expansion are given by
\begin{align}
\mathfrak{L}^{(0)}_{0} & = \sum_{i=0}^{4} f_{i}\dfrac{\dd^{i}}{\dd \xi^{i}} \ ,
\end{align}
where
\begin{align}
\begin{split} 
f_{0} & =315 \left(1536 \, \xi ^2-1536 \, \xi -4175\right) \ ,\\
f_{1} & = 21000 \left(512 \, \xi ^3-768 \, \xi ^2+606 \, \xi -175\right) \ , \\
f_{2} & = 35000 \left(4608 \, \xi ^4-9216 \, \xi ^3+7713 \, \xi ^2-3105 \, \xi +500\right) \ , \\
f_{3} & =  200000 \, \xi  \left(768 \, \xi ^4-1920 \, \xi ^3+1886 \, \xi ^2-909 \, \xi +175\right)\ , \\
f_{4} & = 50000 (\xi -1)^2 \xi ^2 \left(512 \, \xi ^2-512 \, \xi +175\right) \ ,
\end{split} 
\end{align}
and
\begin{align}
\mathfrak{L}^{(0)}_{1} & = \sum_{i=0}^{4} g_{i}\dfrac{\dd^{i}}{\dd \xi^{i}} \ , 
\end{align}
where
\begin{align}
\begin{split} 
g_{0} & = -126 \left(1536 \, \xi ^2-1536 \, \xi -19205\right) \ ,\\
g_{1} & = -8400 \left(512 \, \xi ^3-768 \, \xi ^2+1866 \, \xi -805\right) \ , \\
g_{2} & = -1750 \left(36864 \, \xi ^4-73728 \, \xi ^3+105816 \, \xi ^2-68952 \, \xi +12875\right) \ , \\
g_{3} & =  -10000 \left(6144 \, \xi ^5-15360 \, \xi ^4+17488 \, \xi ^3-10872 \, \xi ^2+850 \, \xi +875\right)\ , \\
g_{4} & = -5000 \, \xi  \left(2048 \, \xi ^5-6144 \, \xi ^4+5524 \, \xi ^3-808 \, \xi ^2-1495 \, \xi +875\right) \ .
\end{split} 
\end{align}

\section{Differential Operators}
\label{app:DifferentialOperators}
The first couple of differential operators in the $ \eta $-expansion of $ \mathfrak{D}^{(4)} $ are
\begin{align}\label{key}
\mathfrak{D}_{-1}^{(4)} &= \frac{1}{\delta_{-1}}\sum_{i=0}^{r}  d_{i}^{(4)}\, \frac{\dd^{i}}{\dd \xi^{i}},
\end{align}
where
\begin{align}
\begin{split} 
\delta_{-1} & = 14400 (\xi -1) \xi  \left(512 \, \xi ^2-512 \, \xi +175\right) \, \\
d_{0}^{(4)} &=  63 \left(1728 \, \xi ^2-1728 \, \xi -29225\right) \ , \\
d_{1}^{(4)} &=  1029000 \left(16 \, \xi ^3-24 \, \xi ^2+18 \, \xi -5\right) \ , \\
d_{2}^{(4)} &=  100 \left(1573184 \, \xi ^4-3146368 \, \xi ^3+2857509 \, \xi ^2-1284325 \, \xi +245000\right) \ , \\
d_{3}^{(4)} &= 4900000 \, \xi  \left(16 \, \xi ^4-40 \, \xi ^3+42 \, \xi ^2-23 \, \xi +5\right)  \ .
\end{split} 
\end{align}
\begin{align}
\mathfrak{D}_{0}^{(4)} &= \dfrac{1}{\delta_{0}}\sum_{i=0}^{r} h_{i}^{(4)} \, \frac{\dd^{i}}{\dd \xi^{i}},
\end{align}
where
\begin{align}
\begin{split} 
\delta_{0}& = 72000 (\xi -1)^2 \xi ^2 \left(512 \, \xi ^2-512 \, \xi +175\right)^2  \ ,\\
h_{0}^{(4)} &= -63 \Big(3735552 \, \xi ^6-11206656 \, \xi ^5-68797824 \, \xi^{4} \\
&\qquad\qquad +156273408 \, \xi ^3-164116480 \, \xi ^2+84112000 \, \xi -25571875\Big)\ ,  \\
h_{1}^{(4)} &=  -840 \Big(2801664 \, \xi ^7-9805824 \, \xi ^6+92446688 \, \xi ^5-206602160 \, \xi ^4  \\
&\qquad\qquad +215959232 \, \xi ^3-122239600 \, \xi ^2+38158750 \, \xi -5359375\Big)\ ,  \\
h_{2}^{(4)} &=  -350 \Big(67239936 \, \xi ^8-268959744 \, \xi ^7+2465438976 \, \xi ^6 \\
&\qquad\qquad\quad -6454957824 \, \xi ^5  +8089782016 \, \xi ^4-5735087360 \, \xi ^3 \\
&\qquad\qquad\qquad\qquad\ \  +2407065875 \, \xi ^2-570521875 \, \xi +61250000\Big) \ ,  \\
h_{3}^{(4)} &=  -2000 \, \xi  \Big(5603328 \, \xi ^8-25214976 \, \xi ^7+296255424 \, \xi ^6 \\
&\qquad\qquad\qquad -919224096 \, \xi ^5 +1320719584 \, \xi ^4-1062573664 \, \xi ^3 \\
&\qquad\qquad\qquad\qquad\qquad +502830650 \, \xi ^2-134474375 \, \xi +16078125\Big) \  .
\end{split} 
\end{align}

\bibliographystyle{utphys}
\bibliography{Refs}

\end{document}